\documentclass[draft]{agujournal2019}
\usepackage{url} 
\usepackage{lineno}
\usepackage{natbib}
\usepackage{soul}
\usepackage{amsmath}

\draftfalse

\journalname{Nature Communications}

\newcommand{\bing}{BING}
\newcommand{\iop}{IOP}
\newcommand{\iops}{IOPs}
\newcommand{\ihop}{IHOP}
\newcommand{\seawifs}{SeaWiFS}
\newcommand{\modis}{MODIS}
\newcommand{\pace}{PACE}
\newcommand{\cdom}{CDOM}

\newcommand{\hydro}{Hydrolight}

\newcommand{\tara}{{\it Tara}}

\newcommand{\emcee}{EMCEE}

\newcommand{\giop}{GIOP}
\newcommand{\gsm}{GSM}
\newcommand{\qssa}{QSSA}

\newcommand{\kparam}{\ensuremath{k}} 
\newcommand{\acst}{\ensuremath{A_{cst}}} 
\newcommand{\aparam}{\ensuremath{A_{exp}}} 
\newcommand{\aphparam}{\ensuremath{A_{ph}}} 
\newcommand{\bparam}{\ensuremath{B_{nw}}} %
\newcommand{\btparam}{\ensuremath{\beta_{nw}}} %
\newcommand{\ston}{\ensuremath{{\rm S/N}}}
\def\kmodel#1{\ensuremath{[\kparam={\rm #1}]}}
\newcommand{\sparam}{\ensuremath{S_{exp}}} 
\newcommand{\spunit}{\ensuremath{{\rm nm^{-1}}}}

\newcommand{\sgiop}{0.018} 
\newcommand{\sgsm}{0.0206} 
\newcommand{\btgsm}{1.0337} 

\newcommand{\nabs}{\ensuremath{a}} 
\newcommand{\abs}{\ensuremath{a(\lambda)}} 
\newcommand{\nbb}{\ensuremath{b_b}} 
\newcommand{\bb}{\ensuremath{b_b(\lambda)}} 
\newcommand{\bbw}{\ensuremath{b_{b,w}(\lambda)}} 
\newcommand{\nbbw}{\ensuremath{b_{b,w}}} 
\newcommand{\bbnw}{\ensuremath{b_{b,nw}(\lambda)}} 
\newcommand{\nbbnw}{\ensuremath{b_{b,nw}}} 
\newcommand{\aw}{\ensuremath{a_{\rm w} (\lambda)}}
\newcommand{\naw}{\ensuremath{a_{\rm w}}}
\newcommand{\anw}{\ensuremath{a_{\rm nw} (\lambda)}}
\newcommand{\nanw}{\ensuremath{a_{\rm nw}}}
\newcommand{\acdom}{\ensuremath{a_{\rm CDOM}}}
\newcommand{\aph}{\ensuremath{a_{\rm ph}(\lambda)}}
\newcommand{\naph}{\ensuremath{a_{\rm ph}}}
\newcommand{\adg}{\ensuremath{a_{\rm dg} (\lambda)}}
\newcommand{\nadg}{\ensuremath{a_{\rm dg}}}
\newcommand{\ag}{\ensuremath{a_{\rm g} (\lambda)}}
\newcommand{\ad}{\ensuremath{a_{\rm d} (\lambda)}}
\newcommand{\nad}{\ensuremath{a_{\rm d}}}
\newcommand{\nag}{\ensuremath{a_{\rm g}}}

\newcommand{\chla}{\ensuremath{Chla}}

\newcommand{\ffaph}{\ensuremath{\naph(440)}} 

\newcommand{\abricaud}{\ensuremath{A_{\rm ph}(\lambda)}}
\newcommand{\bbricaud}{\ensuremath{E_{\rm ph}(\lambda)}}

\newcommand{\kd}{\ensuremath{K_{d}(\lambda)}} 
\newcommand{\rrs}{\ensuremath{r_{rs}(\lambda)}} 
\newcommand{\reflect}{\ensuremath{R_{rs}(\lambda)}} 
\newcommand{\sreflect}{\ensuremath{\sigma(\reflect)}}

\newcommand{\likli}{\ensuremath{\mathcal{L}}}
\newcommand{\dbic}{\ensuremath{\Delta {\rm BIC}}}
\newcommand{\redchi}{\ensuremath{\chi^2_\nu}}
\def\bicij#1{\ensuremath{\dbic_{\rm #1}}}

\newcommand{\grdone}{0.0949}
\newcommand{\grdtwo}{0.0794}
\newcommand{\uparam}{\ensuremath{u(\lambda)}}

\def\smm{\sum\limits}

\newcommand{\pacefile}{ PACE\_OCI.20240413T175656.L2.OC\_AOP.V1\_0\_0.NRT.nc}

\newcommand{\nspec}{3,320} 

\newcommand{\ycdf}{\ensuremath{y_{\rm CDF}}}

\begin{document}

%
%


\title{On the Peril of Inferring Phytoplankton 
Properties from Remote-Sensing Observations}

%
%




\authors{J. Xavier Prochaska\affil{1,2,3,4}\thanks{Current address, Scripps Institution of Oceanography}, Robert J. Frouin\affil{4}}

 \affiliation{1}{Affiliate of the Ocean Sciences Department, University of California, Santa Cruz}
 \affiliation{2}{Department of Astronomy \& Astrophysics, University of California, Santa Cruz}
 \affiliation{3}{Kavli IPMU}
 \affiliation{4}{Scripps Institution of Oceanography, University of California, San Diego}





\correspondingauthor{J. Xavier Prochaska}{jxp@ucsc.edu}





%
%

%
%


\begin{abstract}
Remote-sensing satellites are the only means to observe the entire ocean at high temporal cadence. 
Since 1978, sensors have
provided multi-band images at optical wavelengths to assess ocean color.
In parallel, sophisticated radiative transfer models have been developed
to account for attenuation and emission by the Earth's atmosphere and ocean, 
thereby estimating
the water-leaving radiance or and
remote-sensing reflectance \reflect.
From these \reflect\ measurements,  estimates of the
absorption and scattering by seawater are inferred.
We emphasize an inherent, 
physical degeneracy in the radiative transfer
equation that relates \reflect\ 
to the absorption and backscattering coefficients \abs\ and \bb, known
as inherent optical properties (\iops).
This degeneracy arises because \reflect\ depends on the ratio of 
\nbb\ to \nabs, meaning one cannot retrieve independent functions for 
the non-water IOPs, \anw\ and \bbnw, without a priori knowledge.
Moreover, 
water generally dominates scattering at blue wavelengths and absorption at red wavelengths, further limiting the potential to retrieve IOPs in the presence of noise.
We demonstrate that all 
previous and current multi-spectral
ocean color satellite observations lack the 
statistical power to measure more than three parameters total 
to describe \anw\ and \bbnw.
Due to the ubiquitous exponential-like absorption
by color dissolved organic matter (\cdom), detritus, and 
non-pigmented biomass
at short wavelengths ($\lambda < 500$\,nm),
multi-spectral \reflect\ do not permit the detection of phytoplankton absorption \aph\ 
without very strict priors.
Furthermore, such priors 
lead to biased and uncertain retrievals of \aph.
Hyperspectral observations may recover a
fourth and possibly fifth parameter describing only
one or two aspects of the complexity of \aph.
These results cast doubt on decades of literature 
on \iop\ retrievals, including estimates of phytoplankton growth and 
biomass.
We further conclude that PACE will greatly enhance our ability to measure the phytoplankton biomass of Earth, including its geographic and temporal variations, but challenges remain in resolving the IOPs.
\end{abstract}


%
%

%


%
%
%
%

\section{Introduction}

Phytoplankton play essential roles within our ecosystem, 
serving as the base of the ocean food web and performing
$\sim 50\%$ of all photosynthesis on Earth.
Therefore, 
assessing phytoplankton growth and death -- 
especially in a changing 
climate \citep{behrenfeld2016,flombaum2020}
-- is critical to any effort to track and predict the health
of our planet.
Decades of phytoplankton research have revealed significant regional variations in these process and demonstrated  that phytoplankton are highly dynamic on relatively short time scales (hours to weeks, especially in coastal areas,  due to tides, upwelling, pulses of freshwater inflow,  and other episodic events 
\citep[e.g.][]{Cloern2010}. 
To identify any  long-term trend, therefore, one must first 
develop a detailed picture of the 
variations on seasonal and
shorter timescales. 

Unfortunately, our ability to measure phytoplankton in-situ
is greatly hampered by
the vast expanse of the ocean. 
Measurements with high temporal frequency can only
be acquired at select, fixed stations 
such as OceanSITES \citep{boss2022}.
Therefore, oceanographers have turned to remote-sensing
satellite observations to perform high-cadence,
global analyses of the ocean surface.  
Beginning with the Coastal Zone
Color Scanner experiment \citep{Hovis1980}, 
multi-band observations in 
optical channels have 
enabled the inference of the concentration of 
phytoplankton and other seawater constituents, 
e.g.\   colored dissolved organic matter (\cdom) and
detritus. These inferences are retieved
from satellite-derived remote sensing 
reflectance \reflect, which represents the water-leaving radiance 
normalized by incident solar irradiance.
These are defined and recovered by their absorption 
and backscattering coefficients \abs, \bb, so-called
inherent optical properties (\iops).
  
The underlying physics for \iop\ retrievals 
is radiative transfer:
the absorption and scattering
of sunlight by seawater 
modulates and directs incident sunlight back to the satellite.
While the radiative transfer physics is straightforward
\citep[but not simple;][]{Mobley2022}, 
there are many factors that complicate
the calculations.  These include but are not limited to:
  the concentration of the constituents (typically the desired
  unknown),
  their variation with depth,
  the precise wavelength dependence of the
  absorption and scattering coefficients 
  of each constituent, 
  geometric factors associated with the Sun's location 
  relative to the satellite.
In addition, Earth's atmosphere attenuates the signal
and introduces a dominant background radiation field which
must first be estimated and subtracted (``corrected'')
which fundamentally limits the precision of any 
space-based \reflect\ estimation.

For decades, researchers have attacked this radiative transfer
problem to attempt retrievals of scientifically valuable
quantities including an estimate of the phytoplankton 
biomass. 
There is a robust and well-founded literature describing 
(and performing) the translation of so-called apparent
optical properties (AOPs, e.g.\ \reflect) 
to inherent
optical properties (\iops; \abs, \bb) that depend solely
on the water constituents and the water itself. 
Ideally, one first parameterizes and then estimates 
(``retrieves'') the 
absorption and backscattering spectra of the non-water
component \anw, \bbnw\ and then infers concentrations
of phytoplankton, \cdom, detritus, etc.
From these, one may examine the geographic distribution and
temporal evolution of fundamental biological processes
across the global ocean \citep{fox2022}.  

During the development of a diverse set of \iop\ retrieval
algorithms for this purpose \citep[see][for a review]{werdell2018}, 
the ocean optics community has acknowledged key 
challenges to the problem largely independent of those
associated with radiative transfer.
These include uncertainties related to the atmospheric
corrections, 
non-uniqueness between common constituents (e.g.\ \cdom\ and
detritus), and retrieving multiple unknowns from limited
datasets (e.g.\ multi-spectral observations).
A few, sparsely-cited works have also highlighted a far
more fundamental obstacle to the process: 
a physical ``ambiguity'' in the inversion
of the radiative transfer equation
\citep{sydor2004,defoin2007}. 
Unfortunately, this problem has
often been confused or conflated with the statistical
limitations of an insufficient number 
of bands measuring \reflect\ \citep{werdell2018,cetinic+2024}.
As such, 
while the community has acknowledged challenges to
\iop\ retrievals from remote-sensing observations,
rigorous assessment of the algorithms themselves has
been limited 
and usually only performed in the
context of comparisons to sparse, in-situ observations
\citep[e.g.][]{IOCCG2006,seegers+2018}.

Another fundamental aspect of the problem is that we 
do not know the optimal basis functions that describe
\abs\ and \bb\ nor even the complete set
\citep{garver1994}.
Indeed, it is an aspiration within the ocean color field 
to recover (or even discover) the composition
of phytoplankton 
\citep[e.g.][]{mouw+2017}.
The ocean color research community has hoped that the main limitation
is the sparsity of existing multi-spectral bands provided
by current satellites and that hyperspectral observations will
lead to a major breakthrough.
Indeed, \cite{cael+2023} has demonstrated from a data-driven
analysis of \reflect\ data its limited information
content, i.e.\ only $\sim 2$ degrees-of-freedom
in multi-spectral, satellite observations.
But they also concluded that in-situ hyperspectral datasets
offer only one or two addition degrees of freedom. 
In this manuscript,
we examine this question from a new angle -- with the 
standard approach of \iop\ retrievals -- and reach
similar conclusions.

Here,  we introduce the 
Bayesian INferences with Gordon  coefficients (\bing) 
package for ocean retrievals in a Bayesian context.
Our approach follows many of the standard assumptions
of widely adopted algorithms in the literature, e.g.\ 
the generalized \iop\ (\giop) model \citep{werdell2013},
the Garver-Siegel-Maritorena (\gsm) algorithm
\citep{maritorena2002}.
In addition, we emphasize and expand upon
the ``ambiguity'' problem -- a physical degeneracy
in the radiative transfer equation that
couples reflectances
to \iops\ -- which fundamentally limits \iop\ retrievals.
In turn,  we demonstrate that \iop\ retrievals from
multi-spectral datasets constrain at most three parameters
describing \anw\ and \bbnw.  Therefore, if one allows
for the fact that the spectral shape of \cdom\ absorption
varies and is unknown, then we show one 
cannot retrieve a measurement of phytoplankton absorption.
This includes all previous missions with satellites carrying multi-bands sensors.
We then examine the prospects for \iop\ retrievals
with hyperspectral observations
and discuss additional opportunities to address the 
deep degeneracies that lurk within.

\section{The Bad: A physical degeneracy in the radiative transfer}
\label{sec:methods}

At the heart of our analysis is a new Bayesian inference algorithm for the retrieval of \iops\ 
from remote sensing reflectances, 
the \bing\ package. 
The primary motivations for introducing a Bayesian framework
are twofold: 
 (i) it forces one to explicitly describe all of the priors
 that influence the result;
 and
 (ii) it permits well-established approaches for performing
 model selection, i.e., estimating the maximum number of 
 free parameters one can use to describe the data.
 
To construct any such algorithm, one must have a 
well-defined forward model to predict the observables, 
here remote-sensing reflectances
\reflect.  For \iop\ inversion, this means a radiative
transfer model -- or its approximation -- which
estimates \reflect\ from \abs\ and the backscattering
coefficients \bb.
The majority of \iop\ retrieval algorithms developed
by the community have used the quasi single-scattering
approximation (\qssa) originally introduced by \cite{hansen1971}
and translated to ocean color by \cite{gordon1973}
\citep[see also][]{Zege1991}.
This approach was refined further by \cite{gordon1986} who approximated
the sub-surface remote reflectances \rrs\ 
with a Taylor series expansion: 

\begin{equation}
  \rrs = \smm_{i=1}^{N} G_i \, \uparam^i \;\; ,
\label{eqn:rrs}
\end{equation}
with

\begin{equation}
\uparam \equiv \frac{\bb}{\abs + \bb} \;\; .
\label{eqn:u}
\end{equation}
Most IOP retrieval algorithms have taken $N=2$ and set the
coefficients as constants $G_1 = \grdone$ and
$G_2 = \grdtwo$, i.e.\ values independent of wavelength.
In this manuscript and for the default mode of \bing, 
we adopt the same prescription
and coefficients, but scrutinize the
accuracy of this assumption in Supp~\ref{supp:qssa}.
For the results in the main text, we assume a perfect
forward model, i.e.\ we use
Equation~\ref{eqn:rrs} to generate the target \rrs\ 
and perform the fits on these values. 
In practice, we work with remote-sensing reflectances \reflect\ 
following a standard conversion from \rrs\
\citep{lee+2002}:

\begin{equation}
   \rrs = \frac{\reflect}{0.52 + 1.17\reflect}
\label{eqn:Rrs} 
\end{equation}

For the development and testing of \bing, we have leveraged
a large set of \abs,\bb\ spectra made public by
\cite{loisel23} (hereafter L23).
We use their $X=4,Y=0$ model which includes 
inelastic scattering (not relevant here) and the
Sun at the zenith.
The \iop\ spectra were generated from their database
of in-situ measurements of phytoplankton \aph\ 
and models of several additional constituents: CDOM \ag, 
pure seawater \aw, and detritus \ad.
L23 then generated estimates of the backscattering coefficients \bbnw\ 
following standard assumptions based on in-situ and laboratory work
\citep[see][for additional details]{loisel23}. 
These \nspec\ \abs\ and \bb\ spectra 
define our dataset and range from 350-750\,nm at
$\delta\lambda = 5$\,nm sampling.

\vspace{0.15in}

Despite the approximation of Equation~\ref{eqn:rrs}, 
it does capture a salient aspect
of the physics: the functional dependence of
\reflect\ on \uparam\ and thereby the \iops\ \abs\ and \bb.
However, this dependence is the ``bad'' aspect 
of \iop\ retrievals: because \uparam\ is
a function of the ratio of \bb/\abs:

\begin{equation}
\rrs = Func \left ( \frac{b_b}{a} \right ) \;\; ,
\label{eqn:degen}
\end{equation}
the radiative transfer solutions are 
physically degenerate in \nbb/\nabs.
Put succinctly, any \iop\ solution that recovers
a set of \reflect\ observations can be replaced
by an infinite set that preserves the \nbb/\nabs\ 
ratio.  Therefore, the retrieval is only tractable
if one implements strong constraints 
(known as priors in Bayesian analysis) on the functional
forms of \abs\ and \bb.

In the following section, we examine the consequences
of this physical degeneracy on \iop\ retrievals.
We describe several standard 
priors from previous work and describe their impacts on \iop\ 
retrieval.  Section~\ref{sec:bayes} of the Supplement
provides full details on \bing\ and our specific
approach to perform the Bayesian inference.
We have also implemented standard $\chi^2$ minimization
(Levenberg-Marquardt)
as a fitting option to speed-up model development and
portions of the analysis.
For results in the main text, we assume constant 
signal-to-noise (\ston) for the \reflect\ measurements
unless otherwise specified.

Note that the Bayesian approach in BING involves using Bayes' theorem to update the probability of a hypothesis based on prior knowledge and new data. When retrieving IOPs from \reflect, 
this approach explicitly incorporates all available prior information about the IOPs and their uncertainties into the model. 
By doing so, it allows for a more transparent and rigorous estimation process. The Bayesian framework considers the likelihood of the observed \reflect\ 
given the IOPs and combines it with the prior probability distributions of the IOPs to obtain a posterior distribution. This posterior distribution provides a probabilistic solution to the inverse problem, highlighting the most likely values of the IOPs while quantifying the uncertainties, leading to more reliable and informed retrievals.

\begin{figure}[ht!]
\centering\includegraphics[width=13cm]{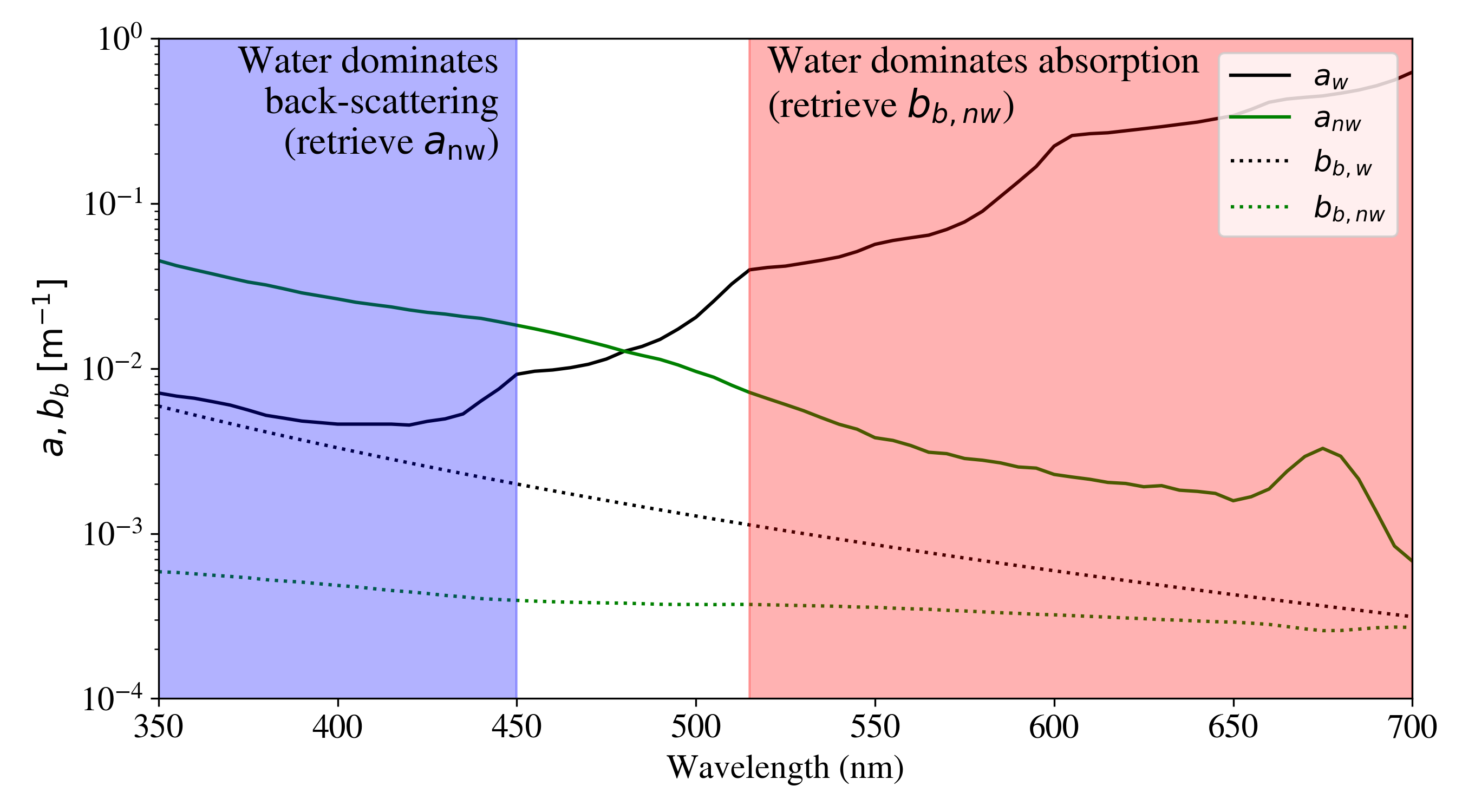}
\caption{Comparison of the \iop\ spectra for water 
 (\naw, \nbbw; black solid and dotted curves) against
 one example set of non-water spectra
 (\nanw, \nbbnw; green solid and dotted curves) from L23 (their index=170).
 The red region indicates where absorption by water dominates
 ($\naw > 5 \nanw$).  In this region, reflectance measurements
 constrain the non-water component of backscattering. 
 Similarly, the blue region is where water dominates backscattering
 and retrievals may constrain \nanw.  In turn, for observations with noise,
 the inversion problem has very limited constraints on the non-water
 components.
}
\label{fig:spectra}
\end{figure}

\section{The Good: Water as a prior}
\label{sec:good}

The ``good'' aspect of \iop\ retrievals is the presence 
of water which introduces an ever-present and precisely
known\footnote{Except in the ultraviolet, 
$\lambda < 400$\,nm, which is intentionally ignored
in this manuscript; \cite{Mason2016}.}
constraint on the problem.
The absorption \aw\ and backscattering \bbw\ spectra of 
pure seawater impose
priors on the model that serve to partially alleviate the physical
degeneracy described in the previous section.
First, \aw\ and \bbw\ span the entire spectrum
and therefore couple the otherwise independent \reflect\ values.
Second, to the extent that the shapes of \naw\ 
and \nbbw\ are unique relative to other constituents
this helps one avoid the \nbb/\nabs\ degeneracy.
Third, the strong absorption of water at 
$\lambda > 500$\,nm and the relatively high 
magnitude of \bbw\ at $\lambda < 450$\,nm
define regions where one may retrieve information on the
non-water components.

On the last point,
Figure~\ref{fig:spectra} compares the absorption
and backscattering coefficients of water against
one example of non-water spectra \anw, \bbnw\ 
from the L23  dataset.  
As emphasized in the Figure,  at red
wavelengths $a \approx \aw$ and 
$\nbbw \approx \nbbnw$ such that  
the observations may constrain \nbbnw.
Similarly at $\lambda < 450$\,nm, 
$\nbb \approx \nbbw$ and $\anw > \aw$
such that the observations constrain \nanw.
These inferences from Figure~\ref{fig:spectra},
however, rely on the strong (but frequently 
satisfied) prior that 
$\aw > \anw$ at $\lambda > 500$\,nm and
$\bbw > \bbnw$ at $\lambda < 450$\,nm.
If this is relaxed, e.g.\ if \anw\ and \bbnw\ 
may take on {\it any} values
then the \nbb/\nabs\ degeneracy forces
an infinite set of solutions (i.e.\ no unique 
retrieval is possible; see Supp~\ref{supp:degenerate}).

Stated another way, 
the greater the freedom that one allows for 
\anw\ or \bbnw, the more degenerate the solutions.
This fundamentally limits our ability to 
retrieve arbitrary \abs\ or \bb\ even in the presence
of {\it perfect} data (infinite number of
channels and no uncertainty).
Therefore, no algorithm\footnote{Including
our own initial efforts with a more sophisticated
algorithm than \bing.} 
can aim to retrieve arbitrary 
\citep{ls1,ls2}
or even highly complex \abs\ and \bb\ 
\citep[e.g.][]{chase+2017}.
To make progress,
one most also impose strong constraints 
on \anw\ and \bbnw\ to recover unique or
most probable solutions.
These priors, however,
must insure that the values of \nabs\ and \nbb\ cannot 
vary at any individual
wavelength where one seeks a retrieval
in a way which holds their ratio constant.

\section{The Ugly: Actual \iop\ retrievals}
\label{sec:ugly}

Let us now perform the ``ugly'': attempted retrievals
of \anw\ and \bbnw\ using assumed spectral
shapes (priors)
with a set of increasingly complex prescriptions.
We follow common practice for models of \anw\ and
\bbnw\ which have been informed by in-situ and
laboratory measurements of ocean constituents.
In turn, we will examine the maximum complexity
that can be
statistically constrained 
by observations 
designed to mimic satellite retrievals, e.g.\ data
from multi-band and hyperspectral observations.

\begin{figure}[ht!]
\centering\includegraphics[width=13cm]{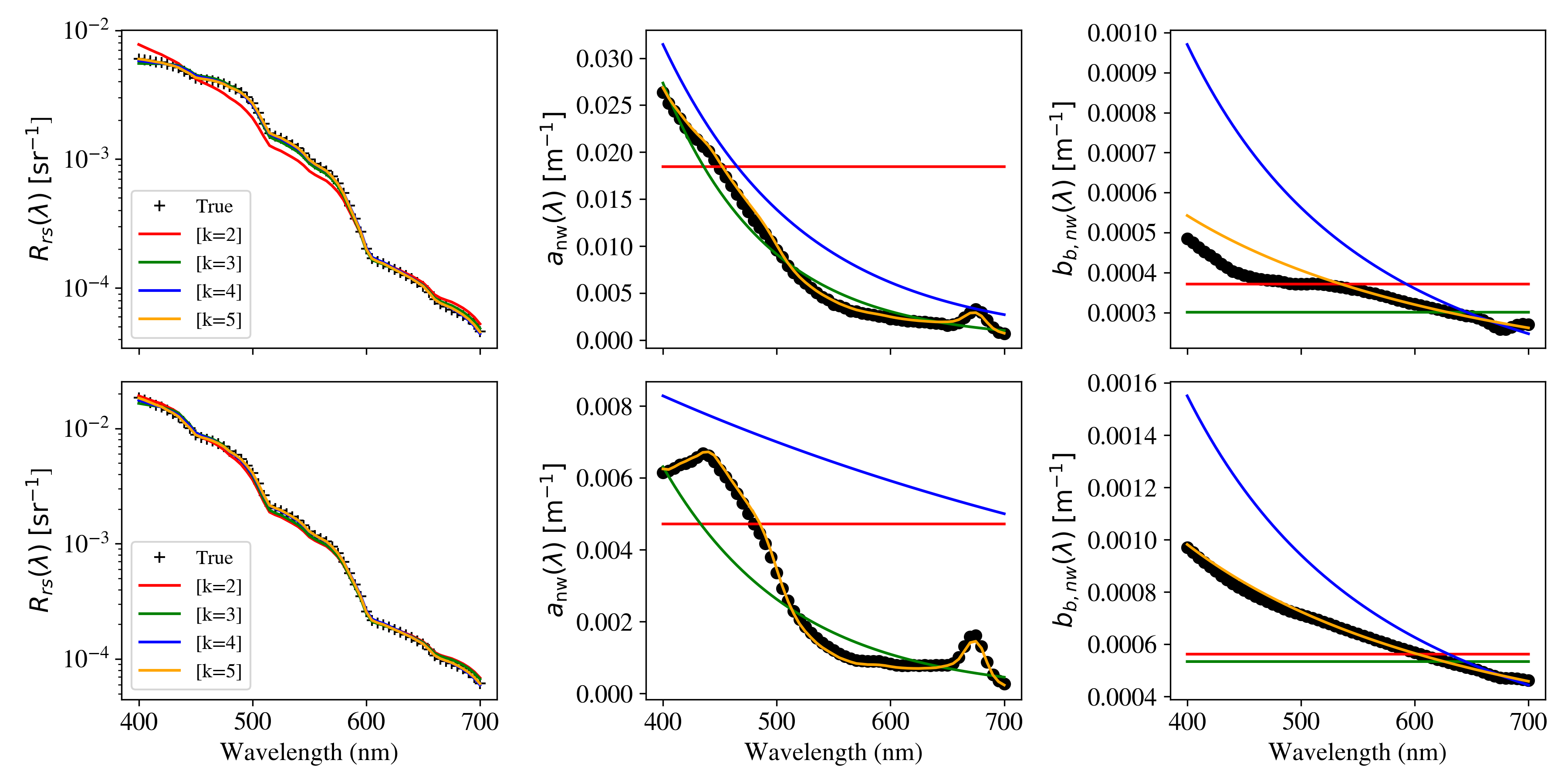}
\caption{Retrievals of \iops\ -- (middle panels) \anw\ and (right panels) \bbnw\ from fits
to (left panels) remote-sensing reflectances \reflect\ assuming perfect radiative
transfer and error-free data.
The black points are the true values of \reflect, \anw, and \bbnw\ for two 
examples from the L23 dataset:
  (index=170) which we chose as representative of the full dataset and
  (index=1032) which has $\naph > \nadg$ at 440\,nm.
The figure shows solutions for a series of models with increasing
complexity and number of free parameters \kparam.
These models are 
    (\kmodel{2}; red) constant \anw\ and \bbnw;
    (\kmodel{3}; green) constant \bbnw\ and a two-parameter exponential for \anw;
    (\kmodel{4}; blue) exponential \anw\ and a power-law for \bbnw;
    and
    (\kmodel{5}; orange) power-law \bbnw\ and \anw\ modeled by the exponential
      and a phytoplankton function (see text for further details).
It is evident that all of 
the $k \ge 3$ models produce excellent fits to the
\reflect\ data at the several percent level.
}
\label{fig:two_fits}
\end{figure}

Consider first the simplest scenario we may conceive:
a two-parameter \kmodel{2} model with both 
\anw\ and \bbnw\ taken as constant at
all wavelengths:

\begin{eqnarray}
 \anw &= \acst \\
    \bbnw &= \bparam 
\end{eqnarray}
We have fitted this model -- including seawater absorption
and backscattering -- to two examples 
from the L23 dataset (Figure~\ref{fig:two_fits}): 
one chosen to be representative
of their full sample and the other chosen to have 
a higher than typical phytoplankton absorption \aph\ 
relative to the combined CDOM and detritus components \adg\ 
(i.e.\ $\naph(440\,{\rm nm}) > \nadg(440\,{\rm nm}$).
For these, we fit to 
\reflect\ values calculated directly from Equation~\ref{eqn:rrs}
and assume a constant \ston\ for the \reflect\ 
values for the likelihood calculation\footnote{The best fits are 
identical for any constant \ston.}.
Despite the extreme simplicity of this \kmodel{2} model,
the \reflect\ fits are not too dissimilar
from the true values, especially for the high
\aph\ example\footnote{This \naph-dominated spectrum
has a low total non-water absorption, i.e.\ weak 
\nadg\ absorption.}.
This follows from our discussion of 
Figure~\ref{fig:spectra}: 
water backscattering and absorption dominates the
solution at 
short and long wavelengths respectively and the non-water
components have limited impact on the \reflect, 
i.e.\ we primarily measure seawater from \iop\ retrievals
especially for ocean waters with low chlorophyll concentrations.

We consider three additional models of increasing complexity:
the \kmodel{3} model,

\begin{eqnarray}
     \anw =& \aparam \, \exp[-\sparam(\lambda-400)] 
     \label{eqn:aexp} \\
    \bbnw =& \bparam 
\end{eqnarray}
the \kmodel{4} model,

\begin{eqnarray}
     \anw =& \aparam \, \exp[-\sparam(\lambda-400)] \\
    \bbnw =& \bparam \, (\lambda/600)^{\btparam} 
\end{eqnarray}
and the \kmodel{5} model,

\begin{eqnarray}
     \anw =& \aparam \, \exp[-\sparam(\lambda-400)] + \aphparam \aph
    \label{eqn:aph} \\
    \bbnw =& \bparam \, (\lambda/600)^{\btparam}
\end{eqnarray}
where $\lambda$ is expressed in nm.
A power-law is frequently adopted to describe \bbnw, especially
that related to particulate matter \citep{gordon1983}.
Here we allow \btparam\ to vary but consider fixed 
exponents in several models discussed in the Supplements.
An exponential function, meanwhile, is commonly used to describe absorption by
CDOM and/or detritus \citep{stramski2001}.
In-situ absorption measurements show typical values of
$\sparam \approx 0.015$ for CDOM \citep{roesler1989} and 
$\sparam \approx 0.010$ for detritus \citep{stramski2001}.
Our fiducial models only require $\sparam > 0$ and
we discuss stricter priors on that parameter below 
(and in Supp~\ref{supp:sexp}).

The \aph\ component in Equation~\ref{eqn:aph} 
is introduced to capture ``typical'' absorption
by phytoplankton.  It is expected and observed that this component may exhibit
the greatest complexity. Indeed, scientifically the community aims
to distinguish the potentially large variations in phytoplankton 
families throughout the ocean and inland waters.
Here, we adopt the parameterization of \cite{bricaud1995}:

\begin{equation}
    \aph = \abricaud \, [\chla]^{\bbricaud}
\end{equation}
with \chla\ the Chlorophyll-a concentration in mg/m$^3$
and the tabulation of
\abricaud\ and \bbricaud\ are provided by \cite{bricaud1998}.
We further set \chla\ to the value used by L23,

\begin{equation}
\chla = \naph(440\,{\rm nm})/ 0.05582 \, \rm m^{-1} \;\; .
\end{equation}
This last step is effectively another prior, but we will
not penalize this model for the
additional information in the analysis that follows.

Figure~\ref{fig:two_fits} shows the best solutions derived with \bing\ 
for the \kmodel{2-5} models.  
While the log-scaling of the \reflect\ panels
hides differences at the few percent level, it 
emphasizes that one requires nearly perfect 
observations to distinguish between the various models.
The \kmodel{3} model matches the data to within 10\%\ 
at all wavelengths and the \kmodel{4} and \kmodel{5} 
models achieve several percent or less.
As a result, the \kmodel{4} model yields a reduced
chi-squared $\redchi \approx 1$ if we assume $\approx 5\%$ errors 
(\ston=20) in the \reflect\ values. 
Visually, at least, Figure~\ref{fig:two_fits} emphasizes the
(ugly) challenge of performing \iop\ retrievals: one is 
severely limited in the degree of unique 
information that can be derived.

\begin{figure}[ht!]
\centering\includegraphics[width=13cm]{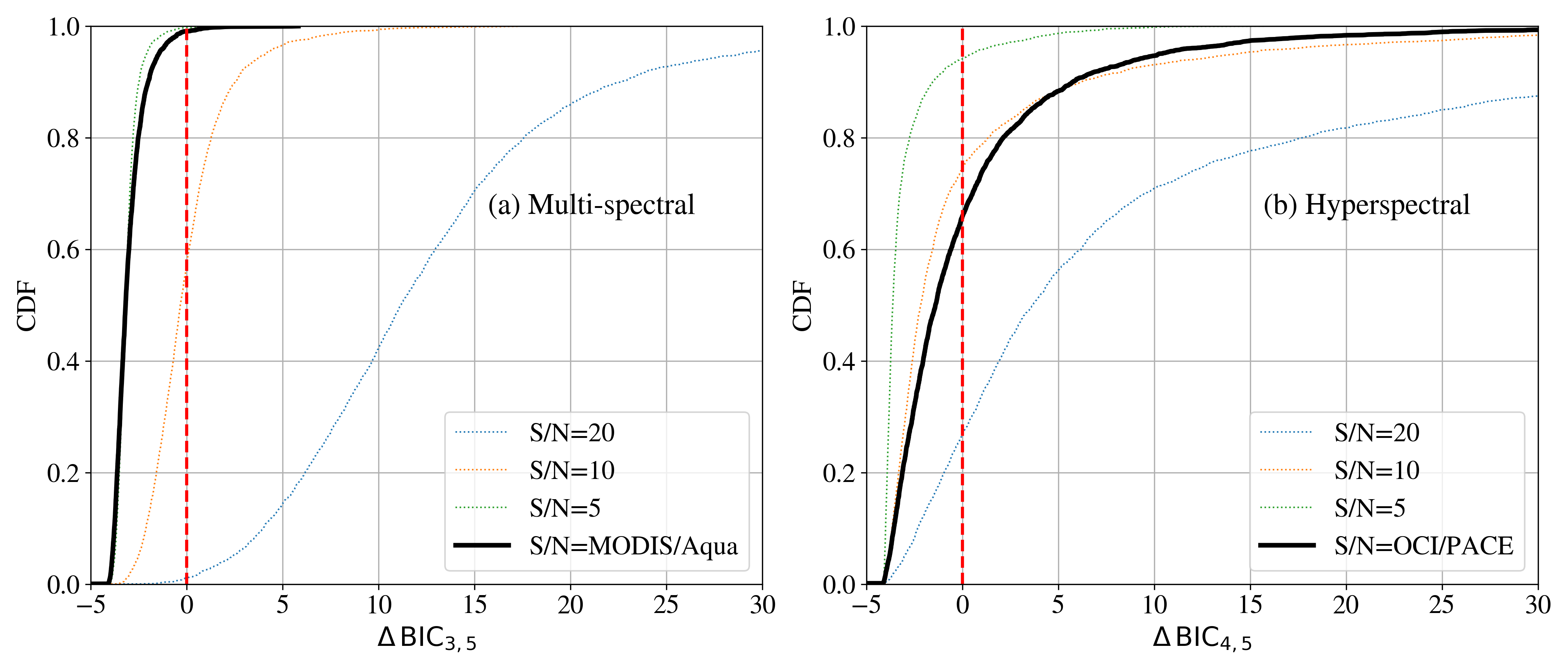}
\caption{
Evaluations of the difference in BIC values \dbic\ for fits to reflectance
data derived from the  \nspec\ spectra of L23.
The curves describe the cumulative distribution function (CDF)
of the \dbic\ values.
The left panel is for simulated \modis\ observations (8~bands)
with a series of signal-to-noise (S/N) 
assumptions (colored curves)
and for retrievals adopting actual \modis\ noise estimates from
NASA validation \cite{seabass}.
The $>99\%$ negative \bicij{3,5} values 
for the realistic noise case indicates 
the L23 spectra greatly favor models with only 3 parameters
and without phytoplankton.
[right panel]  Similar analysis but for OCI/PACE-simulated observations
with fixed \ston\ and for a best guess at nominal OCI/PACE performance
(Supp~\ref{supp:data}).
We find even with OCI that only $\approx 30\%$ of the L23 dataset
favors the model with phytoplankton.
}
\label{fig:BIC}
\end{figure}

We examine these points further in the Bayesian context 
by applying standard approaches for model selection, i.e., 
assessing the balance between model fit and complexity.
Specifically, we have evaluated the 
Aikake and Bayesian information criteria (AIC,BIC)
which are akin to $\chi^2$-difference tests \citep{Bentler1980}: 

\begin{equation}
    {\rm AIC} =  2k - 2 \ln \likli \;\; ,
    \label{eqn:AIC}
\end{equation}
and 
\begin{equation}
    {\rm BIC} =  k \ln n - 2 \ln \likli \;\; ,
    \label{eqn:BIC}
\end{equation}
with $n$ the number of \reflect\ measurements and \likli\ 
the likelihood function 
calculated assuming Gaussian statistics
for uncertainties $\sigma = \reflect/(\ston)$. 
The likelihood function \likli\ 
quantifies the probability of observing the data given the specific forward model and its parameters. Since it is calculated under the assumption of Gaussian (normal) statistics for the 
\reflect\ uncertainties, 
this means that the observed data, i.e., the \reflect\ 
measurements, are assumed to be normally distributed 
around the model predictions with the standard deviation.
For our hyperspectral analysis where $n \gg 10$, the BIC offers
a more stringent constraint but the results with
AIC are qualitatively similar.
A high AIC or BIC value implies that the model is less 
likely to be the best model given the data, 
considering both the fit and complexity.
In the main text, 
we assess model selection by evaluating the difference in BIC
for any two models $\bicij{i,j} = \rm BIC_i - BIC_j$ where
$\bicij{i,j} < 0$ indicates that model~$i$ is preferred and
vice versa.

Figure~\ref{fig:BIC} shows the \dbic\ values
for simulated Moderate Resolution Imaging Spectroradiometer
(MODIS) observations of the L23 spectra 
(see Supp~\ref{supp:data} for details).
The cumulative distribution function (CDF) on the y-axis represents the cumulative probability that the \dbic\ values for the \nspec\ fits to the L23 
\reflect\ data are less than or equal to a specific value. Each \dbic\ value corresponds to a comparison between two models, specifically the BIC values for a model with and without phytoplankton parameters 
(\kmodel{3} and  \kmodel{5} in the multi-spectral case, 
Figure~\ref{fig:BIC}a, and 
\kmodel{4} and \kmodel{5} in the hyper-spectral case, 
Figure~\ref{fig:BIC}b). 
If the CDF value is \ycdf\ at a specific \dbic\ value, 
it means that the $100 \times \ycdf\%$ 
of the \dbic\ values in the dataset are less than or equal to that specific value. Thus, the CDF curve shows the proportion of the dataset for which the simpler model is preferred as a function of the \dbic\ value. 
The higher the CDF value at $\dbic=0$, the higher
fraction of the dataset which
favors the simpler model over the more complex one. 

For the analysis with a realistic MODIS noise model, fewer than 1\%\ of the spectra prefer the \kmodel{5} model with phytoplankton. 
This holds true even though our analysis assumed a perfect forward model and perfect knowledge of the measurement uncertainties, without correlated errors. 
Allowing for these uncertainties would result in zero cases 
with $\dbic < 0$.
In fact, we find (Supp.\ Figure~\ref{fig:BIC_MODIS}) that
one cannot retrieve more than 3 parameters from \modis\ 
observations and that even the \kmodel{2} model is largely
satisfactory.
Without {\it perfect} knowledge of the absorption by \cdom\, 
one cannot retrieve phytoplankton from \modis\ 
observations alone.

The curve corresponding to $\ston = 20$ in 
Figure~\ref{fig:BIC}a shows that while there is some support for the simpler model, indicated by the CDF values for positive \dbic, 
the more complex model, which includes additional parameters for phytoplankton, is generally preferred. This is because the CDF for negative \dbic\ values is low, indicating that the simpler model is not favored in most of the dataset. In other words, although the simpler model is supported in some cases, the overall trend indicates that the more complex model is usually favored for 
$\ston = 20$. Thus, reducing noise in the \reflect\ data is essential when increasing model complexity. However, achieving 
$\ston = 20$ is challenging, even at blue wavelengths in 
open ocean Case 1 waters, as demonstrated in numerous 
validation studies (\ref{supp:data}).
And, achieving $\ston = 20$ at $\lambda > 600$\,nm where
absorption by seawater alone is very high may be impossible
\citep{Zhang2022}.

We reach even stronger conclusions for simulated \seawifs\ observations 
(Supp~\ref{supp:data}) which have fewer bands.
Unless one identifies an approach to regularly achieve
$\ston \gg 10$ measurements in the presence of all error terms
(e.g.\ atmospheric corrections), phytoplankton cannot be retrieved
from multi-spectral observations without strong, 
additional priors.
In Supp~\ref{supp:gg}, we examine the GIOP and GSM models
which assume a fixed and steep \sparam\ shape parameter and the
negative outcomes of this assumption.

Now consider an assessment using L23 spectra with simulated 
Ocean Color Instrument (OCI) 
hyperspectral observations 
on the Plankton, Aerosol, Cloud, ocean Ecosystem
(PACE) sattelite
(see Supp~\ref{supp:data} for 
details).
Our fiducial case uses the L23 spectral sampling
and we limit the observations to 
$400\,{\rm nm} < \lambda < 700$\,nm, outside of which
systematics of the L23 dataset and instrumentation dominate the uncertainties in \reflect,
and poor
knowledge of the wavelength dependence of the ocean's
constituents preclude confident analysis.
Figure~\ref{fig:BIC} shows the distribution in the difference
in BIC values, \dbic, between 
the \kmodel{4,5} models assuming several choices for the
\ston\ and our estimate for the
OCI/PACE noise from v1.0, Level~2 products (Supp~\ref{supp:data}).
We find that OCI/PACE may not recover an assemblage signature
of phytoplankton from water with properties similar to
those represented by the L23 dataset.
We are led to conclude that one can may retrieve 
four parameters for \iops\ from a OCI-like observation
and possibly a fifth.
Two of these numbers describe the amplitude and shape of \anw\ 
 parameterized as an exponential
 and two numbers describe \bbnw\ modeled as a power-law.
Absent strong priors that account for one of these four,
extracting even one number describing \aph\ (at all wavelengths)
will be challenging.

\section{Conclusions and Future Prospects}
\label{sec:future}

In this manuscript, we have introduced \bing, a Bayesian inference
algorithm for \iop\ retrievals utilizing the Gordon coefficients
for radiative transfer.  We have reemphasized a known but 
under-appreciated physical degeneracy in the radiative transfer
--  \rrs\ is a function of \nbb/\nabs\ 
which strictly limits one ability
to retrieve \abs\ and \bb\ without strong priors.
Two of the priors are natural: water both absorbs and scatters light
with precisely known coefficients, at least for wavelengths
$\lambda > 400$\,nm.
We demonstrated, however, that even these constraints
are insufficient; indeed, water frequently dominates the model
limiting the extraction of additional information.
Consequently, we found that
multi-spectral observations with published uncertainties 
\cite{Zhang2022}
cannot reliably retrieve phytoplankton, and that 
even hyperspectral observations (e.g.\ OCI/\pace) will be challenged  
(Figure~\ref{fig:BIC}).  

While the ill-posed nature of the inversion problem to retrieve IOPs from \reflect\ 
is a known issue, this study offers significant value by providing a comprehensive 
analysis and quantification of this degeneracy. The methodology rigorously examines the implications of spectral ambiguities on satellite-derived IOP products, highlighting the specific conditions and parameters where these ambiguities are most pronounced. 
This detailed investigation not only reinforces the necessity for caution in interpreting satellite products, but also advances the understanding of the limitations and potential improvements in current algorithms. By revealing the extent and impact of these ambiguities, 
the study contributes insights for the refinement of remote sensing techniques and the development of more accurate ocean color models. Consequently, this work is a 
crucial step forward in enhancing the precision and reliability 
of ocean color remote sensing.

Previously, \cite{cael+2023} reached a similar inference as
one of our primary conclusions -- the limited information content
of remote-sensing observations.
Specifically, they analyzed the degrees of freedom (DoF) of \reflect\ data
through a standard principal component analysis finding that
in-situ \reflect\ data with MODIS sampling has only 3 DoF
and inferred only DoF=2 for remotely sensed \reflect.
Our analysis, which includes several constraints, such as water
absorption and scattering, yields at least one additional parameter
but the overarching implication is similar:
retrievals from \reflect\ observations have limited information 
content.

On statistical grounds the community could not and cannot
retrieve \aph\ from \reflect\ provided one allows for 
exponential absorption by CDOM and/or detritus 
which are always present.
In fact, this component (\nadg)
tends to exceed \naph, even in the open ocean
\citep{siegel2013,hooker2020}.
Previous work that published estimates of \aph\ 
required very strict priors on the shape
of \sparam\ (Supp~\ref{supp:sexp}), leading to significant
bias in estimates of \naph. 
In the cases where \sparam\ was allowed to 
vary \citep[Boss \& Roesler, Chapter 5][]{IOCCG2006},
the errors on \naph\ were severe and limited retrievals 
only to upper limits on \naph, consistent with this work.

Do our results therefore invalidate the past several decades of 
research and data products using satellite-based ocean color observations?
At the least, all previous retrievals from \reflect\ 
must be further scrutinized. 
We assert uncertainties and biases
were frequently (possibly always) underestimated, and 
substantial correlations between retrieved parameters
will be present. 
Unfortunately even relative analyses of \aph\ may
be subject to large error.
There are, however, key products that are primarily (and very nearly exclusively)
empirical, i.e.\ generated without any radiative transfer model \citep{poc}.
These empirically-based algorithms may circumvent the
radiative transfer issues raised here, but as emphasized
by \cite{cael+2023}, one cannot retrieve an arbitrary 
number of such quantities from visible domain \reflect\ observations.
Therefore, the suite of products
generated by the community to date are 
highly coupled and correlated.
The limited information content of \reflect\ measurements
subject to realistic uncertainties is inherent to the problem.

While the results from Figure~\ref{fig:BIC}b
indicate that hyperspectral observations offer a 
substantial improvement over multi-spectra data,
even {\it detecting} phytoplankton remains challenging.
We reemphasize that the results presented here have assumed
a perfect forward model (i.e.\ no error in the radiative
transfer calculation), uncorrelated uncertainties, a perfect
model for water absorption and backscattering,
and homogeneous seawater (no vertical or horizontal 
spatial variations). 
Furthermore, even if we surmount these issues,
we may only be able to extract a single parameter describing phytoplankton,
e.g., the \naph\ amplitude at $\lambda \approx 440$\,nm.
And, strictly speaking, this may be attributed to {\it any}
absorption component (not solely \naph)
that does not follow the exponential described
by \cdom\ and detritus.

How might we proceed?  
It is abundantly clear that we must identify the optimal
way to parameterize the problem to make most effective
use of the 4 or 5 parameters that describe \anw\ and \bbnw.
For example, if we know \btparam\ (i.e.\ fix its value
as a prior), we would not ``waste'' a free-parameter 
to estimate its value.
In short, we must 
harness our knowledge of the ocean from previous in-situ measurements (or current, if one can afford them) to set priors on the model. These priors should be geographically and temporally variable to reflect different oceanic conditions. 
Because strict and biased priors have been shown to lead to inaccurate and uncertain retrievals,
one must proceed cautiously.
Empirically derived priors can help mitigate these issues by providing a more accurate representation of the ocean's optical properties.

One obvious community-wide effort would be to develop
and agree upon strong priors for \sparam.  This is current
practice in many
existing algorithms (e.g.\ \giop, \gsm, which
set \sparam\ to a single value), but we describe the negative 
consequences of this extreme approach in Supp~\ref{supp:sexp}.
Instead,  we encourage the community to generate \sparam\
priors as probability distribution functions that vary
with geographic location and time 
and then revisit these in our changing climate.
Additionally, we must include more observations, both
from space and in-situ.  From space, we must leverage the
\chla\ fluorescence signal at $\approx 685$\,nm \citep{wolanin2015}
whose production and radiative transfer are distinct from 
that of \iop\ retrievals.
From the ocean, in-situ observations provide invaluable 
validation data and may establish priors like those for 
\sparam. Non-visible in-situ optical observations have been shown to improve retrievals of CDOM absorption, and could be retained to better partition signals related to CDOM and phytoplankton biomass. 
Constraints on \btparam, as a function
of location and season, and physical priors on the
coupling of \abs\ to \bb\ for individual components from the
physics of absorption and scattering could be impactful.

Developing community-wide Bayesian retrieval algorithms is also recommended. The ocean color remote-sensing community should be encouraged to adopt a Bayesian framework for IOP retrievals such as BING, explicitly including all priors and their uncertainties. A Bayesian approach allows for a more transparent and rigorous incorporation of prior knowledge and uncertainties, leading to more reliable retrievals. Unlike BING, however, Bayesian algorithms must adopt an accurate forward model and its uncertainties, include correlated and systematic error in the observations, and harness new data sources. We have initiated such a project – Intensities to Hydrolight Optical Properties (IHOP) – and encourage community adoption and development. The scientists focused on atmospheric corrections have already embarked on this journey following the original insight of 
\cite{frouin2015}. 
Ultimately, we should consider merging the two, i.e., generate an end-to-end Bayesian model that fits top-of-atmosphere radiances to retrieve IOPs.

Increasing the spectral resolution of satellite observations can provide more detailed information about the absorption and backscattering properties of phytoplankton, thereby reducing the impact of degeneracies. Thus, the development and deployment of hyperspectral satellites with high spectral resolution across the visible and near-infrared spectrum are recommended. 
The recently launched PACE satellite will lead the way. 
Additionally, exploring alternative remote-sensing techniques, such as Lidar and fluorescence-based methods, and incorporating polarization information to complement traditional ocean color observations, should be considered. These advanced techniques may provide independent measurements that help resolve ambiguities and improve the overall 
accuracy of phytoplankton estimates, although information gained using these new techniques should be clearly demonstrated and defined first in the field.

Addressing the uncertainty in atmospheric corrections is another critical recommendation. Improving atmospheric correction algorithms to reduce the uncertainties they introduce into ocean color retrievals is crucial. Incorporating advanced atmospheric models and ancillary data (e.g., aerosol properties) can enhance correction accuracy. Atmospheric corrections are a significant source of error in remote-sensing reflectance measurements, and reducing these uncertainties is essential for accurate phytoplankton IOP retrievals.
Continued implementation of dark-pixel correction schemes, for example, removes non-visible information and is inconsistent with in-situ observations of the aquatic light field.

Promoting interdisciplinary collaboration is also essential. Fostering collaboration between oceanographers, remote-sensing experts, and radiative transfer modelers to address the complex challenges of IOP retrievals can bring together diverse expertise and perspectives, leading to more innovative and effective solutions.
These should include individuals with mastery of statistics
who can rigorously assess uncertainty and help develop
robust and transparent algorithms.

By implementing these recommendations, the remote-sensing community can significantly enhance the accuracy and reliability of phytoplankton IOP retrievals, leading to better-informed biogeochemical models and ecological assessments. This comprehensive approach will help ensure that remote-sensing data accurately reflect the true state of the ocean's biological and chemical processes, thereby supporting more effective environmental monitoring and management efforts.

\section{Supplementary Materials}

\begin{table*}
\centering
\caption{MODIS Data \label{tab:modis}}
\begin{tabular}{cc}
\hline 
Band & \sreflect \\ 
(nm) & (sr$^{-1}$) \\ 
\hline 
412 & 0.0012 \\ 
443 & 0.0009 \\ 
488 & 0.0008 \\ 
531 & 0.0007 \\ 
547 & 0.0007 \\ 
555 & 0.0007 \\ 
667 & 0.0002 \\ 
678 & 0.0001 \\ 
\hline 
\end{tabular} 
\\ 
Notes: The error has assumed that 1/2 of the variance is due to the in the in-situ measurements. \\ 
\end{table*} 

\begin{table*}
\centering
\caption{SeaWiFS Data \label{tab:seawifs}}
\begin{tabular}{cc}
\hline 
Band & \sreflect \\ 
(nm) & (sr$^{-1}$) \\ 
\hline 
412 & 0.0014 \\ 
443 & 0.0011 \\ 
490 & 0.0009 \\ 
510 & 0.0006 \\ 
555 & 0.0007 \\ 
670 & 0.0003 \\ 
\hline 
\end{tabular} 
\\ 
Notes: The error has assumed that 1/2 of the variance is due to the in the in-situ measurements. \\ 
\end{table*}

\begin{figure}[ht!]
\centering\includegraphics[width=13cm]{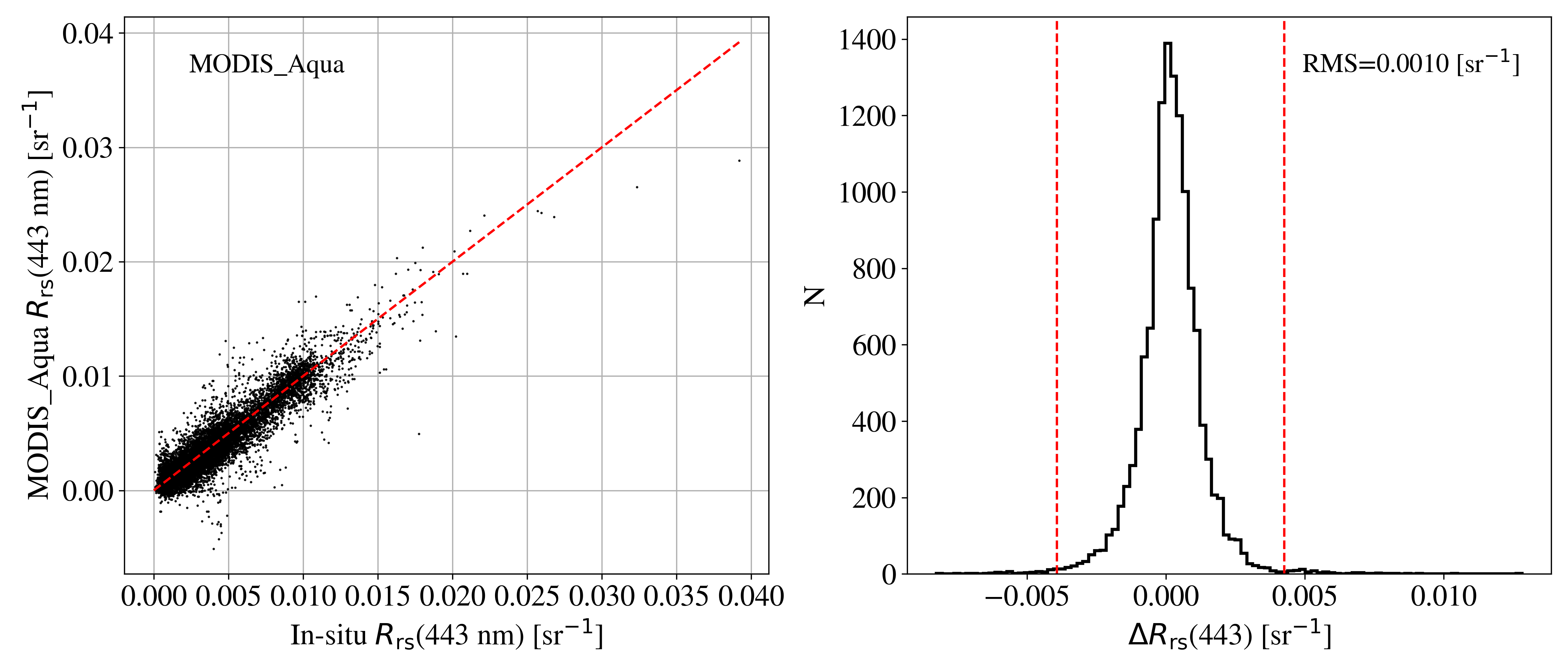}
\caption{(left) Comparison of the \modis\ measurements
of \reflect\ at $\lambda = 443$\,nm against
in-situ observations at the same wavelength.  These
were taken from the SeaBASS site \citep{seabass} dedicated to \modis\ 
matchups \citep{modis_matchup}.
The points follow the over-plotted one-to-one line
relatively well, albeit with significant scatter which
we assess as the RMS noise in the \modis\ observations.
(right) Distribution of the difference between in-situ
and satellite $\Delta \reflect = \reflect^{\rm in-situ}-\reflect^{\rm MODIS}$.
The red dashed-lines show the $4\sigma$ interval beyond
which we clipped the data when calculating the noise
estimate (RMS).
}
\label{fig:modis_noise}
\end{figure}

\begin{figure}[ht!]
\centering\includegraphics[width=13cm]{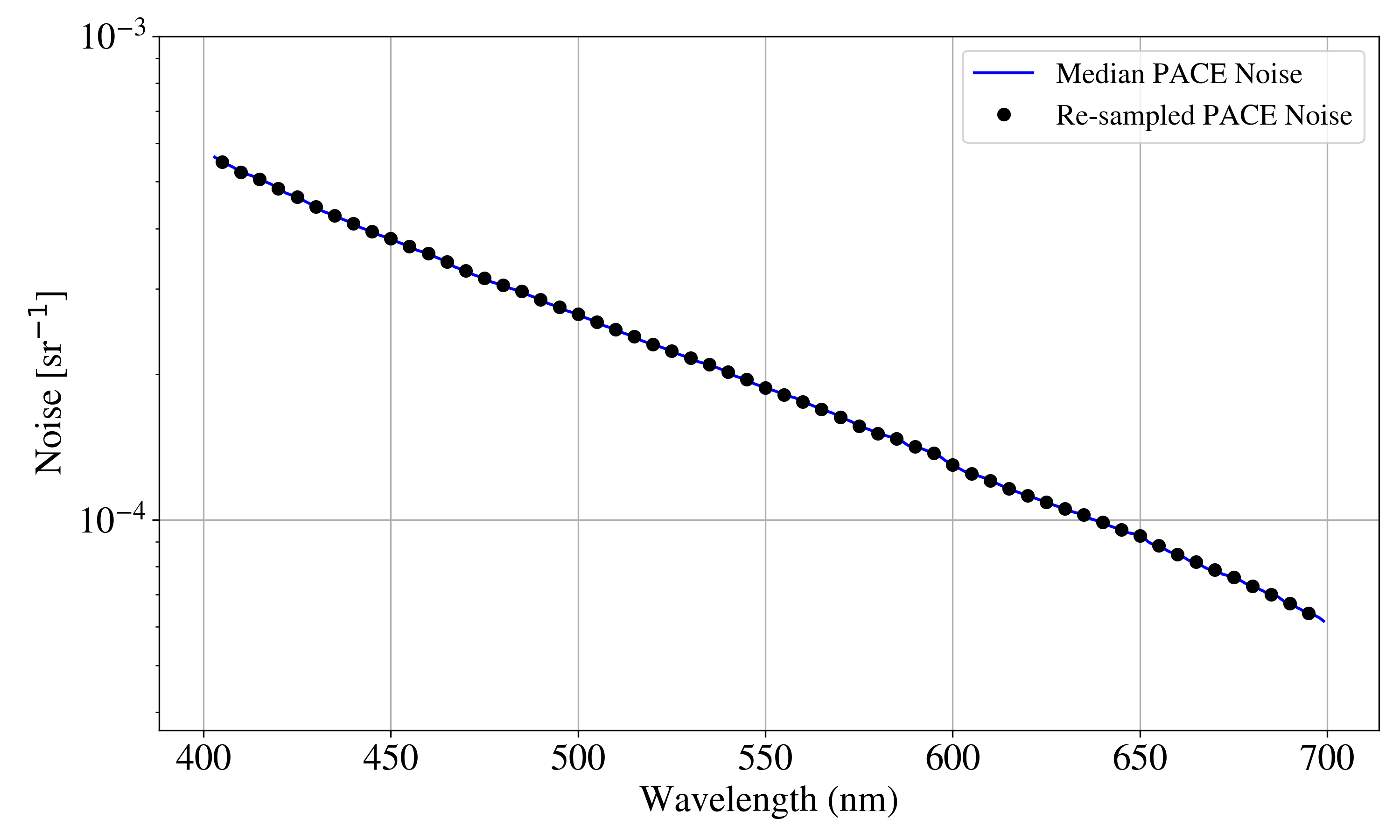}
\caption{The blue curve is the median PACE uncertainty in \reflect\ 
estimated in the Level~2 product, v1.0 for one granule (\pacefile).
The black dots are the values of \sreflect\ adopted in this manuscript 
when simulating \pace\ spectra with noise.
}
\label{fig:pace_noise}
\end{figure}

\subsection{Data and their Uncertainties}
\label{supp:data}

As described in the main text, all of the absorption and scattering spectra
examined in this manuscript were taken from \cite{loisel23} (aka, L23).
Their dataset includes total absorption and scattering as well as
the contributions from individual components (e.g.\ \naph, \nad, \nag).  
The dataset also 
includes outputs from \hydro\ calculations \citep{hydrolight} using these spectra as
input, e.g.\ \reflect\ and depth-resolved quantities like the
diffuse absorption coefficient \kd.
Specifically we use the L23 dataset with inelasctic scattering 
and at 0\,deg. solar and viewing zenith angle.

For many of the analyses presented here, we have independently
generated simulated \reflect\ spectra 
for several multi-spectral and hyperspectral missions.
For the fits presented in this manuscript, we 
ignore the \reflect\ provided by L23 and
instead use Equations~\ref{eqn:rrs}-\ref{eqn:Rrs} to calculate
\reflect\ from \abs\ and \bb.
We then resample these spectra to the bands/channels of
several satellite missions:

\noindent
\underline{MODIS/Aqua}: We adopt 8 multi-spectral bands as listed
in Table~\ref{tab:modis}
corresponding to the MODIS Aqua mission and
we evaluate \reflect\ at the center of each. 
For uncertainties,  we have estimated the RMS difference
between satellite and in-situ \reflect\ ``match-up'' measurements
collated on the SeaBASS database \citep{seabass} 
after iteratively clipping any $4\sigma$
outliers.  Figure~\ref{fig:modis_noise} shows an example
of the data and clipping for one band.
We further assumed that one half of the variance is due
to the in-situ observations themselves.
These RMS values are also provided in Table~\ref{tab:modis},
and we find they are in good agreement with other estimations
\citep{Zhang2022,kudela2019}.

\noindent
\underline{\seawifs/SeaStar}: We followed a similar procedure for 
the Sea-viewing Wide Field-of-view Sensor (\seawifs)
using 6 bands and the uncertainties provided
in Table~\ref{tab:seawifs}.

\noindent
\underline{OCI/\pace}: For simulated OCI/\pace\ spectra, we assumed 
$\delta=5$\,nm  sampling and limited the wavelength
range from $400-700$\,nm.   The lower bound is due
to 
  (i) greater uncertainties in the atmospheric corrections,
  (ii) greater uncertainty in water absorption and scattering,
  (iii) greater uncertainty in how best to parameterize 
   the non-water components in the UV.
The upper wavelength bound is to avoid systematics that
likely dominate the uncertainty at the lowest \reflect\ signals.

For the \pace\ noise model, we downloaded a single 
granule (2,175,120~pixels)
of Level~2 data, v1.0: \pacefile.
We then took the median uncertainty spectrum 
\citep[Rrs\_unc][]{Zhang2022}
for all non-flagged data
between 33-40N and 73-78W.  This median spectrum is plotted
in Figure~\ref{fig:pace_noise} at the Level~2 wavelength
sampling ($\approx 2.5$\,nm).
We also show the values adopted at our $\delta = 5$\,nm
sampling, and one notes that we did not adjust \sreflect\ 
despite the larger sampling size.
This is because OCI is over-sampled at $\delta=2.5$\,nm,
i.e.\ the data is highly correlated.
Further work should assess the degree of this correlation
to obtain a better noise estimate.
If we were to assume no correlation
(i.e.\ reduce \sreflect\ by $1/\sqrt{2}$), 
the results in Figure~\ref{fig:BIC}b 
would yield a higher fraction of spectra
preferring 5~parameters:
instead of $\approx 70\%$ of the L23 dataset showing 
$\dbic<0$, we find $\approx 30\%$ fail the criterion.

\subsection{Taylor expansion model re-examined}
\label{supp:qssa}

\begin{figure}[ht!]
\centering\includegraphics[width=13cm]{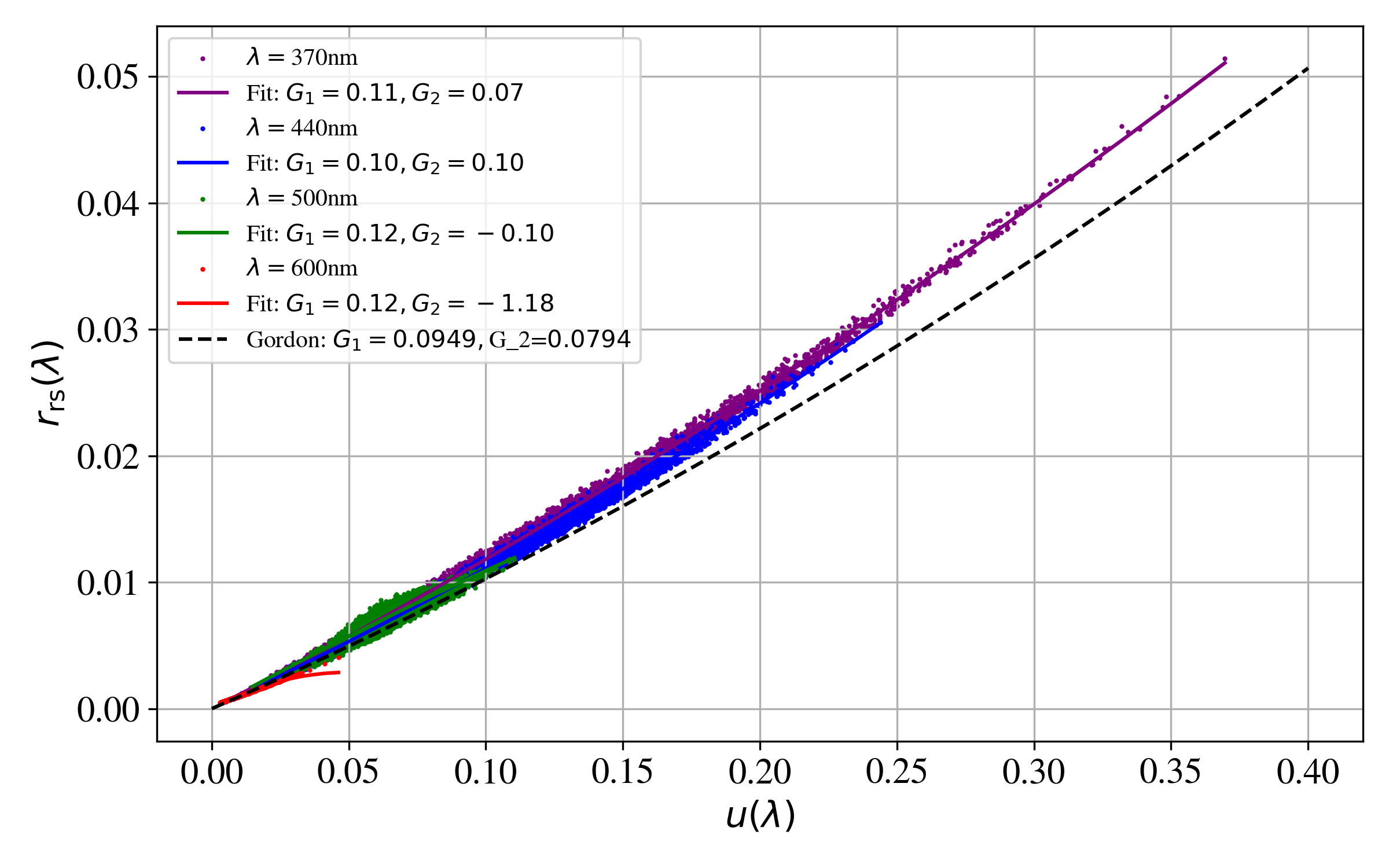}
\caption{Sub-surface reflectances \rrs\ generated with the
\hydro\ radiative transfer code by L23
(converted from \reflect\ using Equation~\ref{eqn:Rrs})
against \uparam\ as defined by Equation~\ref{eqn:u}.
The black dashed line shows the second order Taylor expansion
of \rrs\ in \uparam\ (Equation~\ref{eqn:rrs}) with the 
Gordon coefficients most widely adopted by the community.
In general, this curve underpredicts \rrs\ as calculated
with \hydro\ with a maximum offset of $\approx 10\%$ at $\uparam = 0.35$
and $\lambda=370$\,nm.  
We also show a series of individual fits
of Equation~\ref{eqn:rrs} to the data with the legend indicating
the derived $G_1, G_2$ coefficients.
Note that $G_1$ is largely independent of wavelength
but that $G_2$ is strongly wavelength
dependent (and anti-correlated).
}
\label{fig:u}
\end{figure}

To examine the use of Equation~\ref{eqn:rrs} as
an approximation of the radiative transfer,
consider Figure~\ref{fig:u} which plots 
the \hydro-derived \reflect\ of L23 converted
to \rrs\ with Equation~\ref{eqn:rrs} against
the evaluated \uparam\ values using
Equation~\ref{eqn:u} at 4 distinct wavelengths.
These evaluations approximately follow a quadratic
with zero y-intercept.
Overplotted with a dashed line is the Gordon
approximation using the standard $G_1,G_2$ 
coefficients and Equation~\ref{eqn:rrs}. 
Qualitatively, the \hydro\ outputs follow the
relation yet lie systematically above the curve.
At its extreme, the Taylor series approximation is offset by
$\approx 10\%$ at $\lambda = 370$\,nm and 
$\uparam=0.35$.  

To further illustrate the difference, we have 
fitted $G_1, G_2$ coefficients to the data 
at select wavelengths and recover similar $G_1$
values but $G_2$ values that vary significantly
with wavelength 
 ($G_2 \approx 0.07$ at $\lambda = 370$\,nm, 
 $G_2 \approx -1.2$ at $\lambda = 600$\,nm). 
We also find that there is significant
scatter around each of the fits with a relative
RMS of $\approx 5\%$ at shorter wavelengths and $20\%$ at
the reddest wavelengths.
We expect this scatter is inherent to Equation~\ref{eqn:rrs}
and would be unavoidable if one uses this approximation
even with wavelength-dependent coefficients.
An accurate retrieval algorithm would need to 
account for these variations or otherwise 
suffer significant systematic error.  This
is the focus of a separate algorithm we are
developing (\ihop), and we also refer the readers
to recent advances in approximations of the radiative
transfer equation \citep{twardowski2018}.

\subsection{Bayesian Inference}
\label{sec:bayes}

Provided the forward model and a parameterization of \abs\ and \bb,
the Bayesian inference is straightforward and a wealth of well-trodden
approaches and software packages are available.
For \bing, we adopt 
the Monte Carlo Markov Chain (MCMC) formalism
which empirically derives the posterior probabilities for
the \abs, \bb\ parameterization $P(X|Y)$ including
full uncertainties and all of the cross-correlation terms.
This requires the definition of a likelihood function
$P(Y|X)$, 
which will have the form:

\begin{equation}
P(Y|X) \propto \exp \left \{ -\frac{1}{2} [Y-H(X)]^{\rm T} C [Y-H(X)] \right \}
\end{equation}
where $Y$ represents the measured \reflect\ values,
$C$ is the full covariance matrix of \reflect\ 
including correlations,
and $H(X)$ is the forward model of \reflect\ at the locations of $Y$. 

It can be shown that an MCMC analysis converges to the exact
solution if run for an infinitely long time;
in practice, the calculations tend to converge after 
$\approx 10,000$ iterations.  For the analysis here, we generally run
for 75,000~trials with at least 2 walkers per parameter
(and at least 16~walkers) and only analyze the last
7,000~iterations of each. 
This release of \bing\ uses the \emcee\ sampler
\citep{emcee} which was developed for astrophysical applications
and has see wide-spread adoption in the field
(over 8,000 citations).

\begin{figure}[ht!]
\centering\includegraphics[width=13cm]{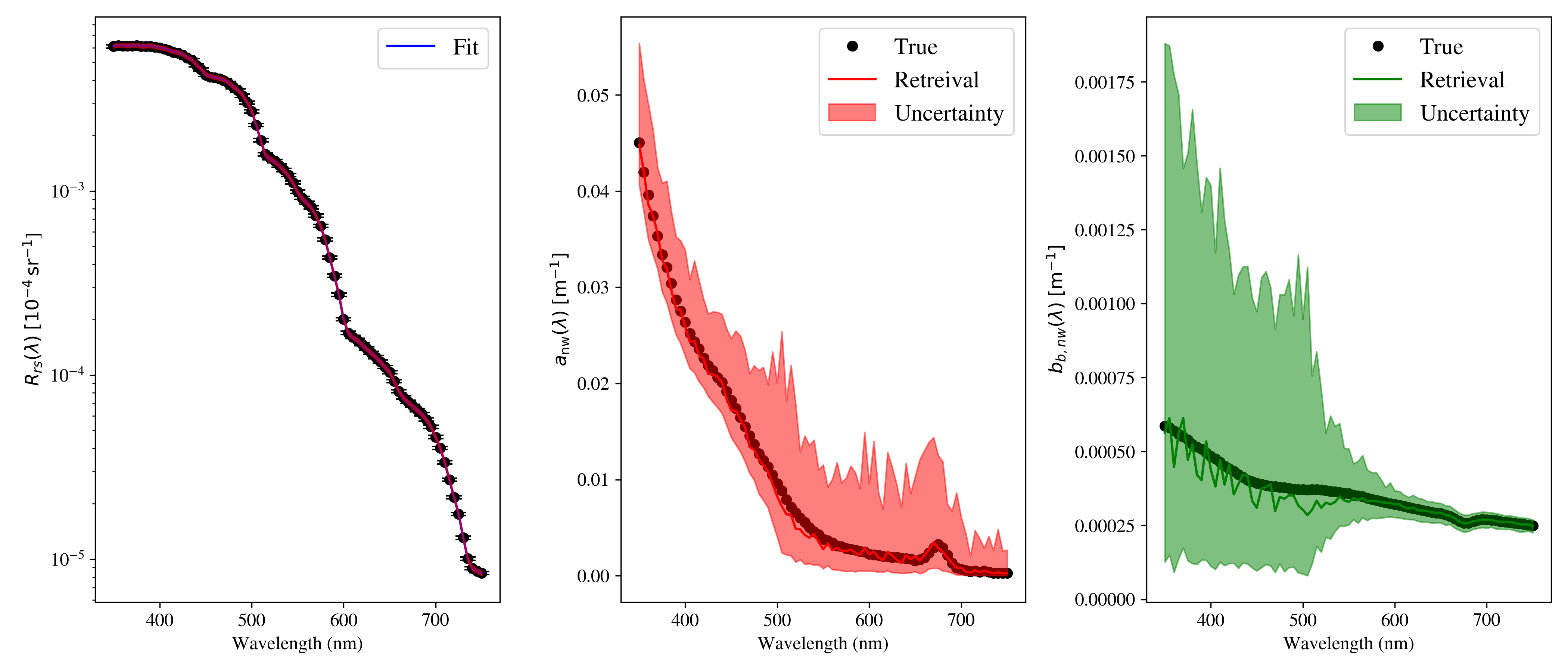}
\caption{Example of a \bing\ retrieval of \anw\ (center) and \bbnw\ (right)
from the \reflect\ spectrum (left).  In this example, we parameterized
\anw\ and \bbnw\ with 81~free parameters, one for every wavelength channel.
Owing to the physical degeneracy in the radiative transfer equation,
$\reflect = F(a/b_b)$, there are an infinite number of solutions yielding
an infinite uncertainty in \anw\ and \bbnw.
Here, the Bayesian inference was only performed for 70,000 steps and has
not fully described the (infinite) posterior space (and never would).
}
\label{fig:degen1}
\end{figure}

\begin{figure}[ht!]
\centering\includegraphics[width=13cm]{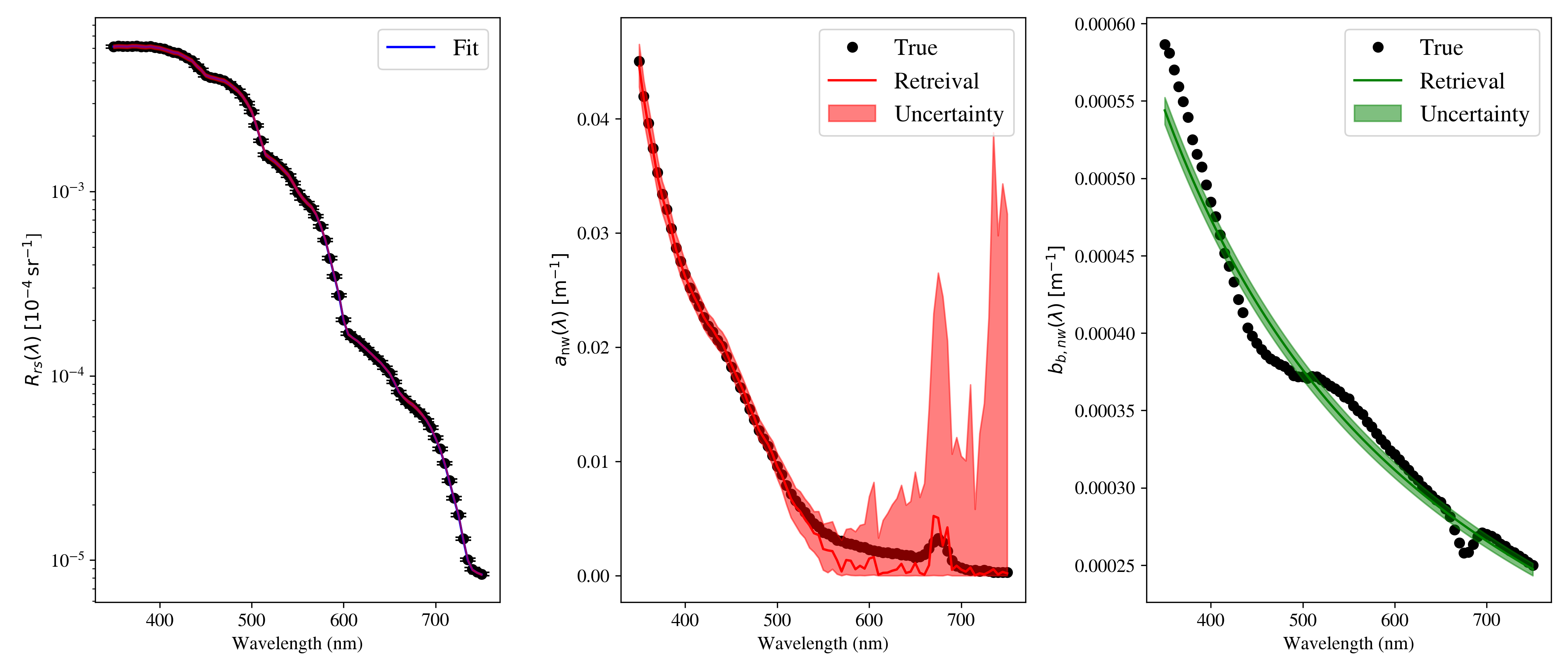}
\caption{
Same as Figure~\ref{fig:degen1} but with \bbnw\ 
parameterized as a power-law with fixed exponent and 
\anw\ still allowed to have any shape.
Again, there are an infinite number of solutions, only 
partially captured by the 75,000 steps of the
sample.
}
\label{fig:degen2}
\end{figure}

\subsection{Fits with \abs\ and/or \bb\ free to vary}
\label{supp:degenerate}

One point emphasized in this manuscript is that one cannot retrieve
arbitrary \abs\ or \bb\ or even \anw\ and \bbnw\ because of a 
physical degeneracy in the inversion of
the radiative transfer equation (Equation~\ref{eqn:degen}).
To demonstrate this with two examples, we performed an \iop\ retrieval
of \anw\ and \bbnw\ assuming a perfect forward model (Equation~\ref{eqn:rrs}),
perfect knowledge of the uncertainties, and perfect knowledge of water
(\naw, \nbbw).  
In this case, we solve for ``arbitrary'' \anw\ and \bbnw\ 
by parameterizing each with 81~free parameters, one
per wavelength channel.
We show the fits to \reflect\ for the index=170
spectra of L23 in Figure~\ref{fig:degen1} and the retrievals with 
95\%\ confidence intervals, finding that our solutions
for \anw\ and \bbnw\ start ``blowing up" after only 50,000 samples.

To further emphasize the point, we refit the data but constrained 
\bbnw\ to be a power-law with a fixed exponent
\btparam\ set at a value that closely corresponds
to the known \bbnw\ spectrum and only 
allow the amplitude \bparam\ 
of \bbnw\ to vary in the retrieval.  
Once again, the solution diverges (Figure~\ref{fig:degen2})
although more slowly.  If we were to run for infinite samplings,
we would recover an infinite uncertainty.
With an arbitrary \anw\ spectrum, we can match any change 
to the amplitude of \bbnw\ and maintain constant
$a/b_b$.  Therefore, even with the strong constraint
of water and just a single degree of freedom for
\bbnw, we still cannot retrieve arbitrary \anw.
This physical degeneracy limits the information 
content of retrievals and precludes arbitrary functional
forms for \anw\ and \bbnw, e.g.\ models 
which strive to retrieve arbitrary
\anw\ are ruled out.

\begin{figure}[ht!]
\centering\includegraphics[width=13cm]{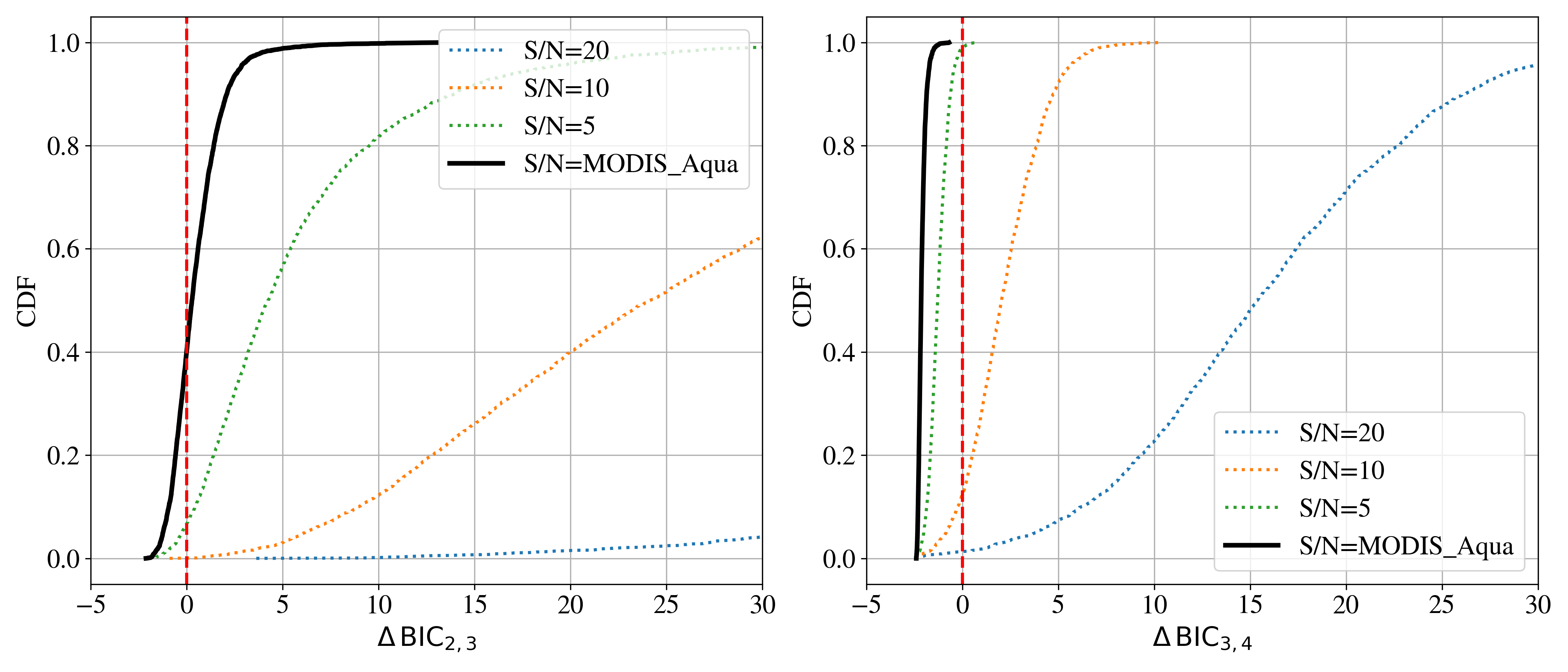}
\caption{Additional BIC analyses on simulated \modis\ spectra.
The (left) panel compares the \kmodel{2} and \kmodel{3} models where we find
that only $\approx 50\%$ of the L23 dataset favors 
anything more complex than the two-parameter,
constant model.
The (right) panel demonstrates that the \kmodel{4} model is disfavored for
the entire L23 dataset.
}
\label{fig:BIC_MODIS}
\end{figure}

\begin{figure}[ht!]
\centering\includegraphics[width=13cm]{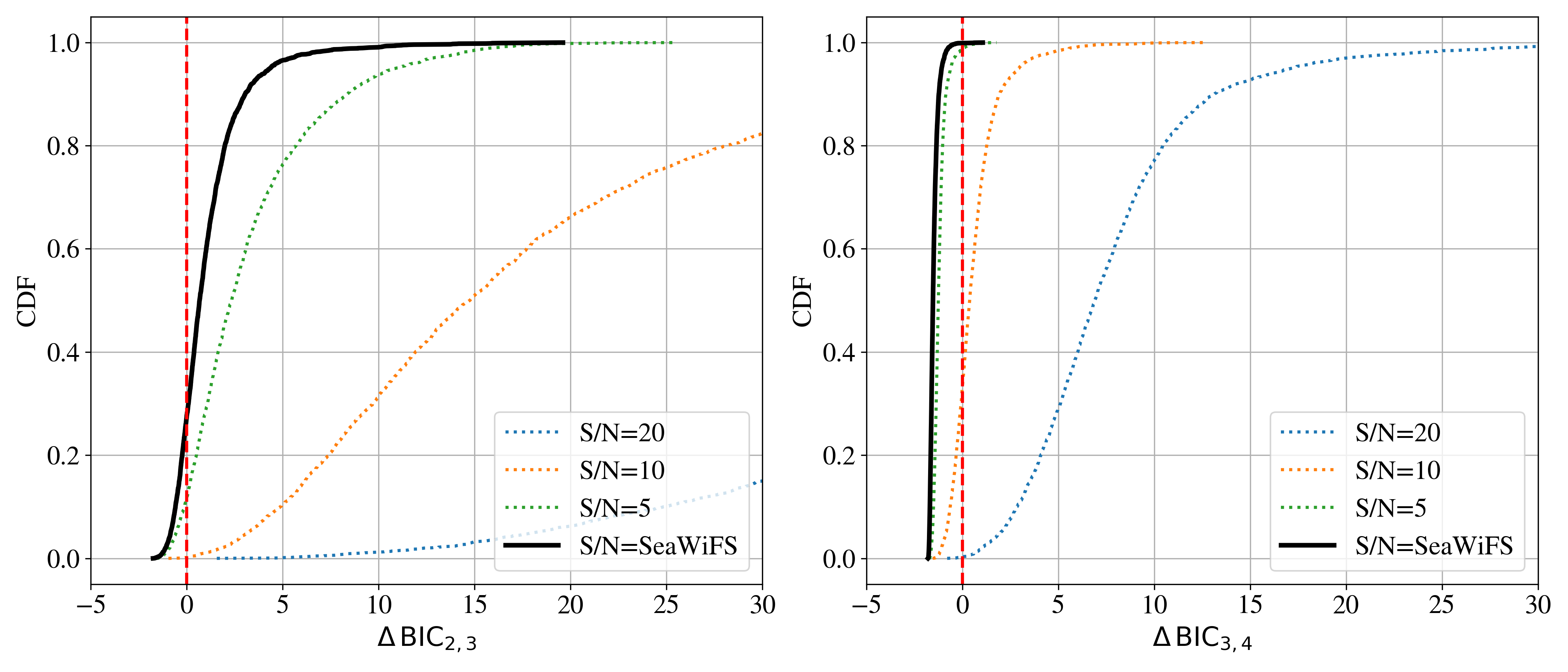}
\caption{BIC analysis of simulated \seawifs\ spectra
for (left) the \kmodel{2} vs.\ \kmodel{3} models
and (right) \kmodel{3} vs. \kmodel{4}.
Approximately 30\%\ of the L23 dataset favors the 
\kmodel{2} model and only a few favor the \kmodel{4} model.
Similar to \modis, multi-spectra observations generally
prefer a model with $\kparam<3$ parameters.
}
\label{fig:BIC_SeaWiFS}
\end{figure}

\begin{figure}[ht!]
\centering\includegraphics[width=13cm]{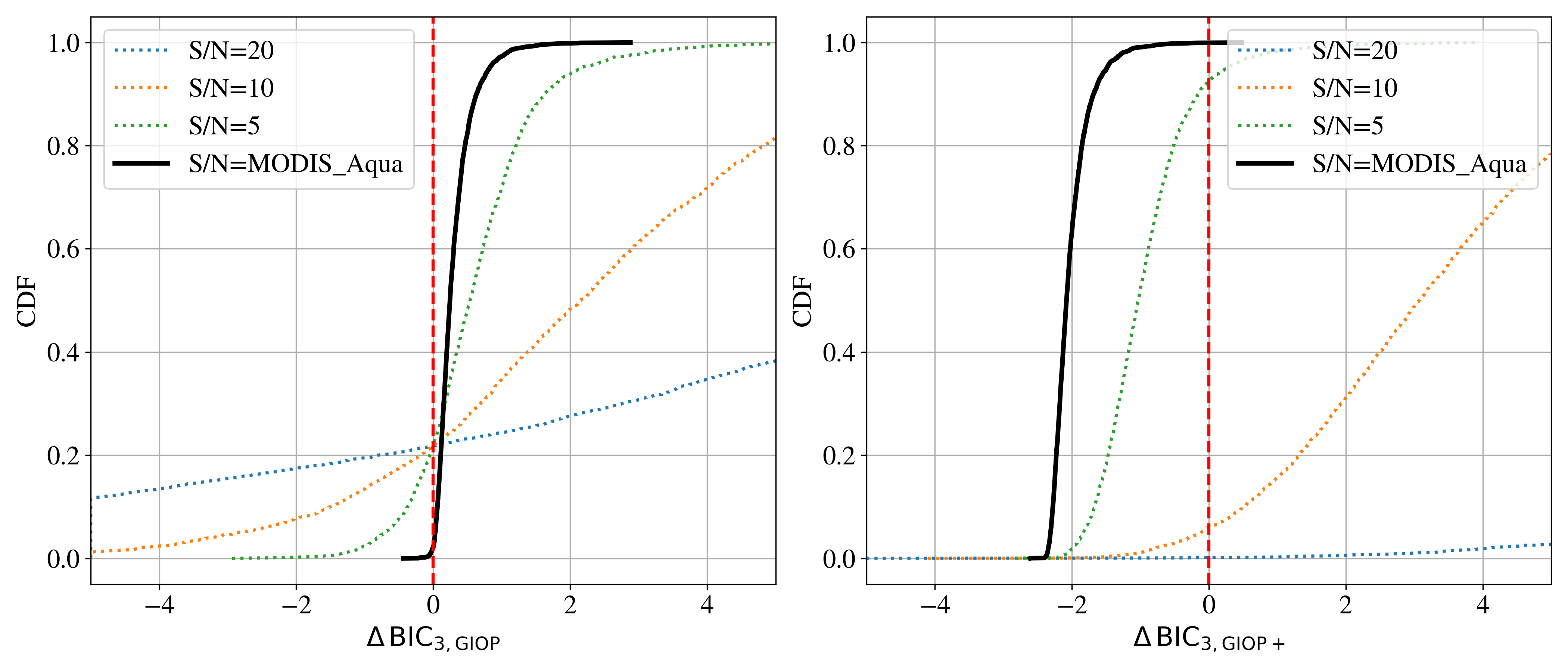}
\caption{These panels describe the BIC analysis assuming MODIS-like
observations for two models designed to match standard GIOP configurations.
The (left) panel compares our \kmodel{3} model against 
the standard \kparam=3 parameter
configuration of GIOP (see text for full details).
We find this GIOP model is preferred, which we speculate is due to the 
extra freedom to fit the \reflect\ at blue wavelengths where the \ston\ 
in \modis\ is highest.  
(right) Results for the GIOP+ model (\kparam=4 parameters) which
lets \btparam\ be an additional free parameter.
This model is disfavored for the entire dataset, further evidence
that one cannot recover 4~parameters from \modis\ observations.

}
\label{fig:BIC_GIOP}
\end{figure}

\begin{figure}[ht!]
\centering\includegraphics[width=13cm]{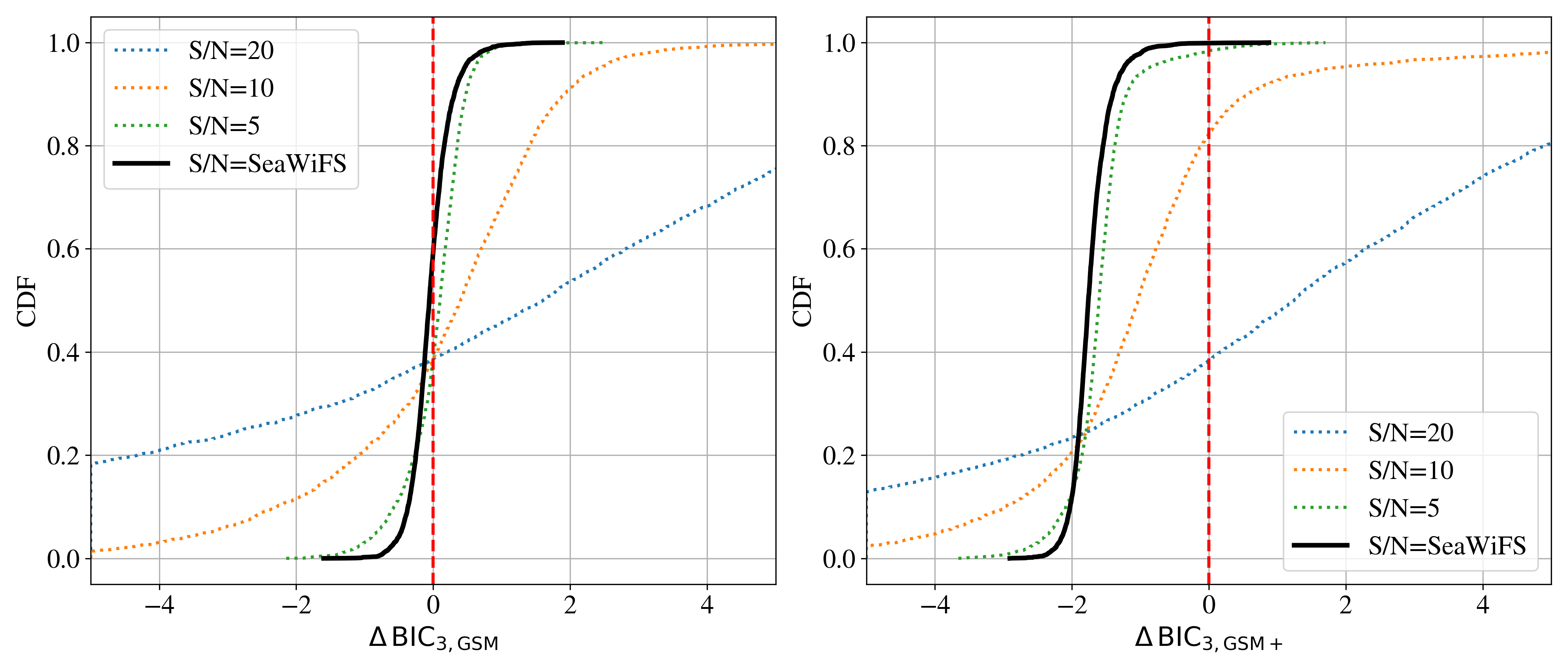}
\caption{Similar to Figure~\ref{fig:BIC_GIOP} but for
GSM models and using simulated \seawifs\ spectra.
As with the GIOP models, we find the GSM model is favored
over our \kmodel{3} model but that a \kparam=4 parameter
version -- GSM+ which lets \btparam\ be free --
is highly disfavored.
}
\label{fig:BIC_GSM}
\end{figure}

\subsection{Additional BIC Analysis}
\label{supp:mo_bic}

We have compared additional pairs of models for the 
simulated multi-spectral \reflect\ spectra than
those in the main text.  These additional analyses
explore further the conclusion that such datasets
are limited to retrieving a total of \kparam=3 parameters
describing \anw\ and \bbnw\ from multi-spectral,
satellite observations.

Figure~\ref{fig:BIC_MODIS} shows \dbic\ distributions
for the \kmodel{2} vs.\ \kmodel{3} model and the 
\kmodel{3} vs.\ \kmodel{4} model. Remarkably, we
find that approximately half of the L23 dataset if 
optimally described by the constant \iop\ model.
The results in Figure~\ref{fig:BIC_MODIS} show 
clearly that the \kmodel{4} model is disfavored
for {\it all} of the L23 dataset.
This latter result greatly supports the conclusion
on multi-spectral data given in
the main text.
Not surprisingly, the conclusions are even stronger
for \seawifs\ spectra (Figure~\ref{fig:BIC_SeaWiFS}).

Now consider Figure~\ref{fig:BIC_GIOP} which presents
\dbic\ distributions for two ``GIOP'' models which assume
an exponential component for \anw\ with fixed shape
parameter \sparam=\sgiop\spunit,  
the Bricaud formulation
for \aph\ and either (i) GIOP: 
the \cite{Lee2002} 
approach to estimating the power-law exponent of \bbnw\
or (ii) GIOP+: allowing \btparam\ to be free.
This means \kparam=3 free-parameters for GIOP\footnote{
Although we note that the shape of \aph\ and \bbnw\ 
are informed by the data such that it may
be considered to have greater than \kparam=3.
}
--
  \aparam, \aphparam, and \bparam --
and a \kparam=4 for GIOP+ (\btparam). 
Figure~\ref{fig:BIC_GIOP} shows
that this model is favored over our
fiducial \kmodel{3} model which has no
phytoplankton but free \sparam.
This implies that strong priors on \sparam\ can
lead to the inference of additional absorption
(e.g.\ phytoplankton).  We show below, however, the
negative consequences of this overly strict prior,
especially adopting a steeper slope than that typically
found in the ocean.
The figure also demonstrates that another parameter
(GIOP+)  cannot be extracted from the data.

Figure~\ref{fig:BIC_GSM} shows \dbic\ distributions
for two ``GSM'' models which also have a fixed 
exponential but with an even steeper slope
(\sparam=\sgsm\spunit) and a power-law for \bbnw\ with
fixed exponent \btparam=\btgsm\ (GSM) or letting
\btparam\ be free (GSM+).
Therefore, these are also \kparam=3 and \kparam=4 models 
with the same parameters as GIOP and GIOP+.
We find that this mode is favored over the \kmodel{3}
model for only 33\%\ of the L23 dataset.
Similar to the GIOP model, we show in Supp~\ref{supp:sexp}
that its \aph\ 
retrievals are both highly biased and uncertain.



\begin{figure}[ht!]
\centering\includegraphics[width=13cm]{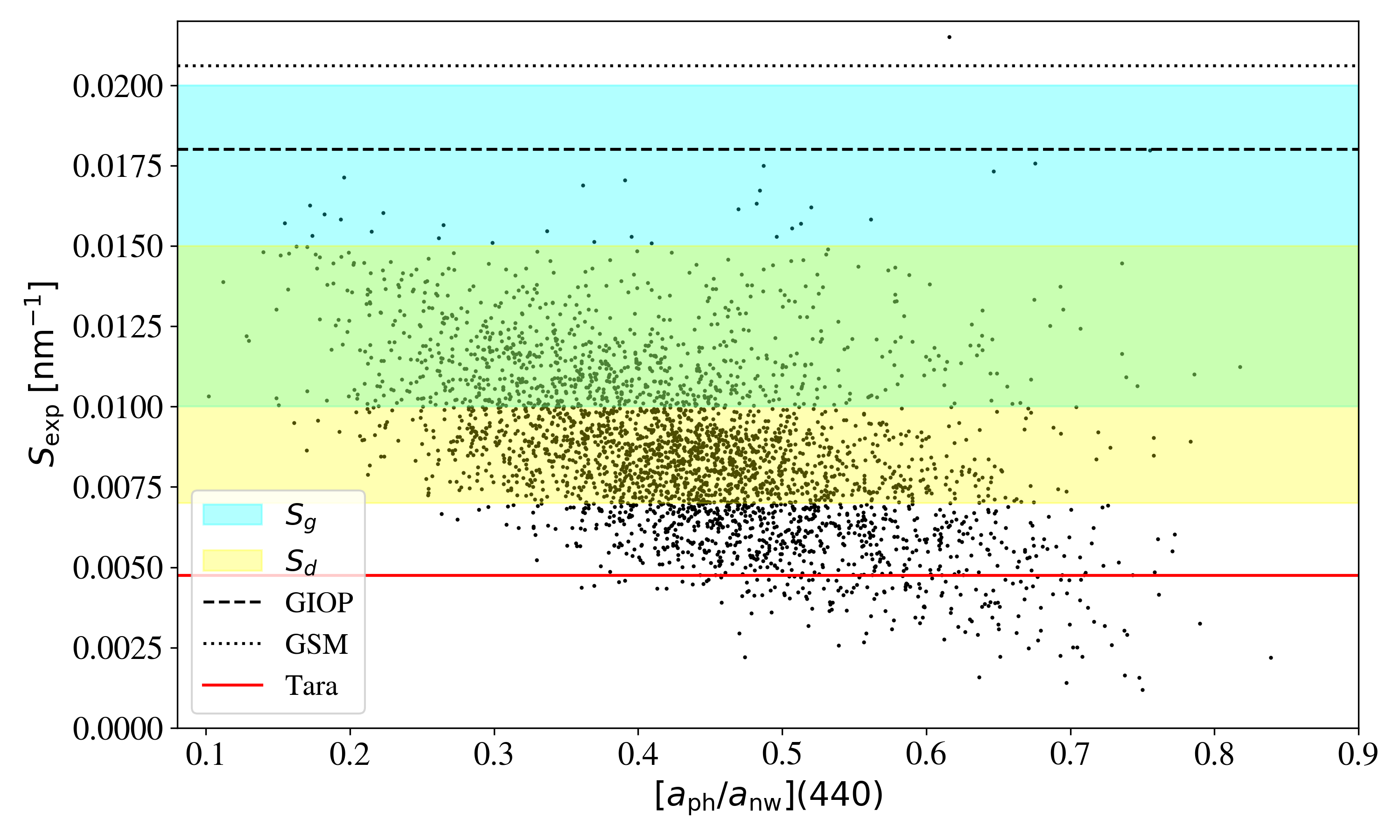}
\caption{The dots plot the best fit shape parameter \sparam\ 
for the L23 spectra using the \kmodel{4} model (no phytoplankton
component) versus the amplitude of \naph\ to \nanw\ at 440\,nm.
The two are correlated, albeit with large scatter.
The blue and yellow shaded regions indicate the ranges
of exponential shapes for \cdom\ and detritus 
($S_g, S_d$) respectively adopted by L23.
Not surprisingly, the majority of retrieved \sparam\ lie
within these loci.
The ones with shallower slope, however, may be attributed
to the presence of phytoplankton which effectively flattens
\anw\ at $\lambda \approx 450$\,nm.
The black dotted/dashed lines demarcate the fixed values
of \sparam\ assumed by the \giop/\gsm\ algorithms for \adg.
These are steeper values than the typical $S_g$ (and all $S_d$)
values adopted by L23.  In addition, we shows the \sparam\ 
derived from an extreme, \cdom-subtracted \tara\ absorption
spectrum collected off the coast of Africa 
\citep[see][]{pg2024}
which demonstrates at least one instance of a very low 
\sparam\ in the ocean.
}
\label{fig:Sexp}
\end{figure}

\subsection{Treatment of \sparam}
\label{supp:sexp}

The previous section introduced the GSM and GIOP models
which adopt fixed \sparam\ shape parameters for the
exponential term in \anw.
We now examine that choice in the context of the 
L23 dataset and then consider the consequences
for \aph\ retrievals in the following sub-section.

For our fiducial treatment of the exponential component 
in \anw\ (Equation~\ref{eqn:aexp}), the only constraint placed
on the shape parameter \sparam\ was that it be positive-definite
($\sparam > 0 \, \rm nm^{-1}$).
Let us scrutinize that prior as it affects
the potential to retrieve phytoplankton and any other
constituents (see the previous section). 
Figure~\ref{fig:Sexp} shows the \sparam\ values derived
with the \kmodel{4} model (no phytoplankton) against the 
fraction of non-water absorption associated 
with phytoplankton at 440\,nm in the L23 spectra,
\naph/\nanw.
The two quantities are anti-correlated because the increased
presence of phytoplankton relative to \cdom\ and detritus
tends to give a shallower, non-water absorption spectrum.

We find, as anticipated, that the 
majority of retrieved \sparam\ values
lie within the loci of shape parameters assumed by L23 
for \cdom\ and detritus based on \citep{IOCCG2006}.
There is, however, a non-negligible set of retrieved
\sparam\ values that are lower than the lowest
value assumed by L23;  these are partially due
to strong phytoplankton
absorption.  If one were to adopt a stricter prior 
on \sparam\ than $\sparam > 0\,\rm nm^{-1}$, 
therefore, one may find statistical
evidence for phytoplankton absorption
(formally, a distinct component from the 
best-fit exponential).
Indeed, we found this to be the case for the GIOP and
GSM models (Figures~\ref{fig:BIC_GIOP},\ref{fig:BIC_GSM})
although it is now apparent from Figure~\ref{fig:Sexp}
that their choices for \sparam\ lie at the upper end
of \sparam\ values measured in the ocean (GIOP)
or even beyond (GSM).
This was by intentional for GSM \citep{maritorena2005},
as its designers derived fixed values for \sparam\ and \btparam\ 
by achieving best retrievals when compared against in-situ data.
However, their treatment of errors may not have been
well suited to actual \seawifs\ uncertainties
(Table~\ref{tab:seawifs}) and we show below that the
model is both biased and yields highly uncertain
\aph\ estimates (in fact, primarily upper limits).




\begin{figure}[ht!]
\centering\includegraphics[width=13cm]{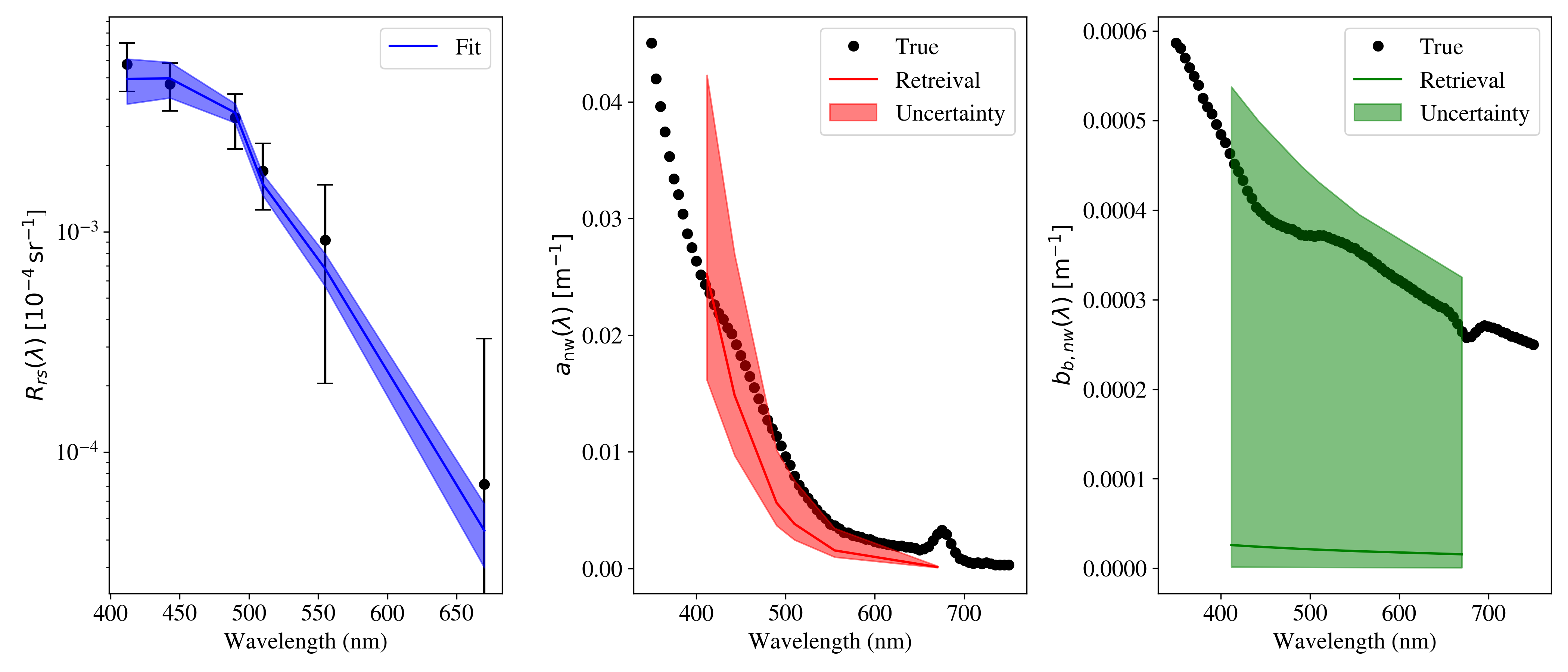}
\caption{\bing\ inferences from the (left) 
\reflect\ spectra for (center) \anw\ and 
(right) \bbnw.  These results are for a fit
to the index=170 spectrum of L23 assuming
\modis\ observations and the GIOP model.
Note that we have not perturbed the data by 
the estimated \modis\ noise but instead evaluate
at their actual values.
The shaded regions show 
90\%\ confidence intervals from the posterior
draws.  Clearly, the \bbnw\ and \anw\ estimations
are highly uncertain.
}
\label{fig:GIOP_fit}
\end{figure}

\begin{figure}[ht!]
\centering\includegraphics[width=13cm]{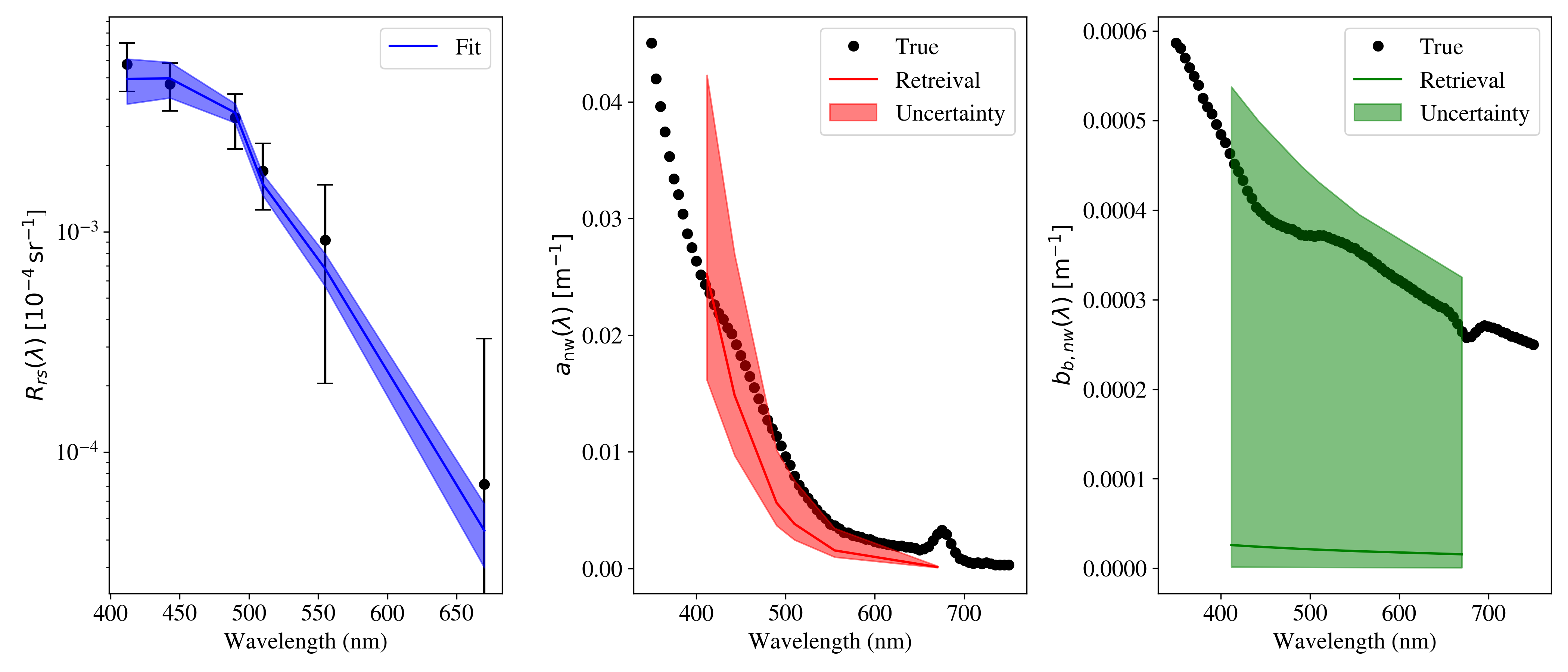}
\caption{Same as Figure~\ref{fig:GIOP_fit}
but for GSM fitted to simulated \seawifs\ 
\reflect\ spectra.
}
\label{fig:GSM_fit}
\end{figure}

\begin{figure}[ht!]
\centering\includegraphics[width=13cm]{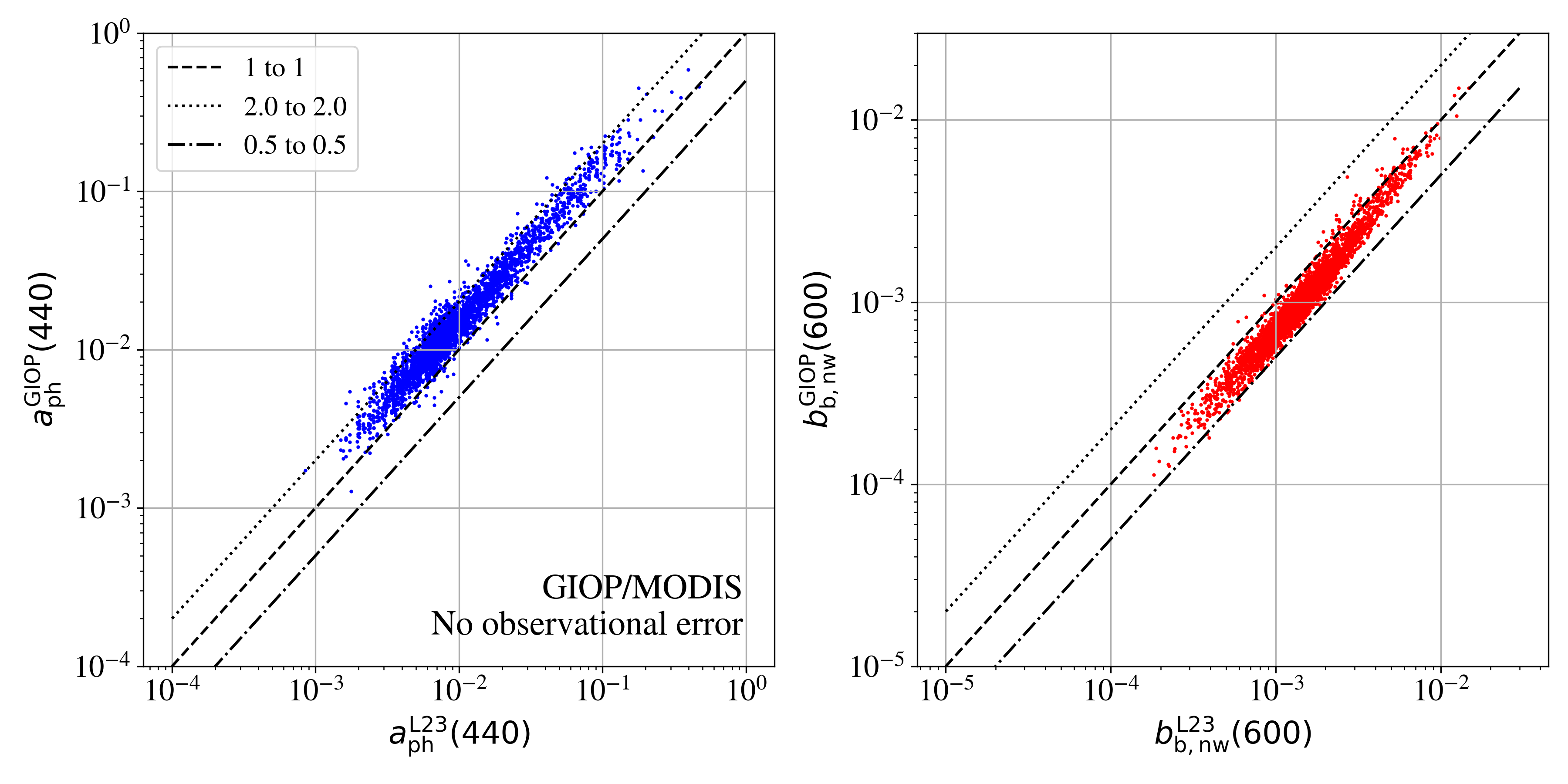}
\caption{Comparison of retrieved phytoplankton
absorption (left) and particulate backscattering
(right) from GIOP on simulated MODIS spectra
against the known values of the L23 dataset.
These are for a perfect forward model and zero
error. It is evident that the \aph\ values
are biased high, by approximately 50\%.
Similarly the \bbnw\ values from GIOP (at 600\,nm)
are low by $50\%$ at small values and $\sim 20\%$
at larger values.
}
\label{fig:GIOP_perfect}
\end{figure}

\begin{figure}[ht!]
\centering\includegraphics[width=13cm]{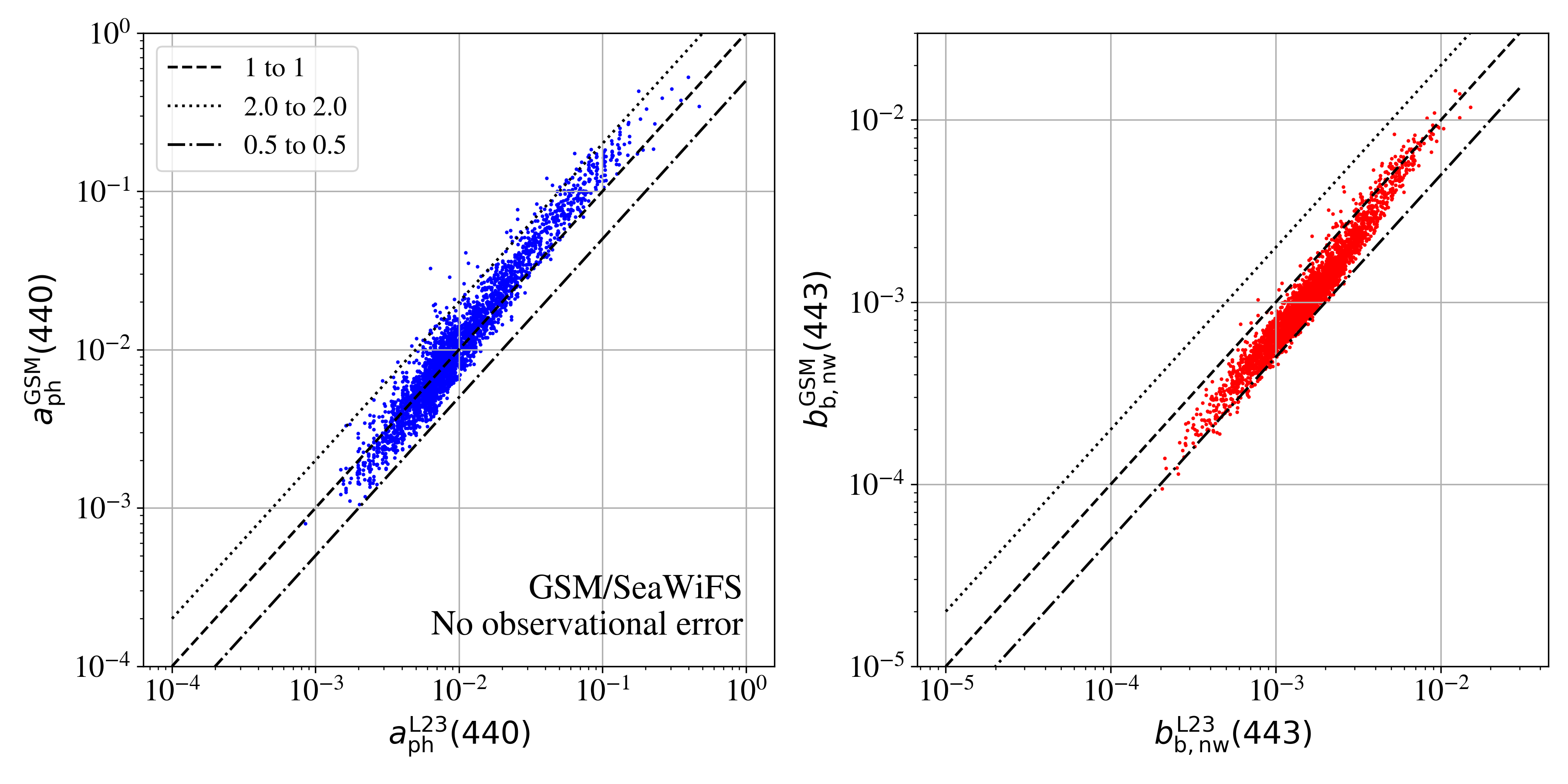}
\caption{Same as Figure~\ref{fig:GIOP_perfect}
but for the GSM model and simulated \seawifs\ 
spectra.  As for GIOP, we find significant
biases in the retrievals of several to many
tens of percent.
}
\label{fig:GSM_perfect}
\end{figure}

\begin{figure}[ht!]
\centering\includegraphics[width=13cm]{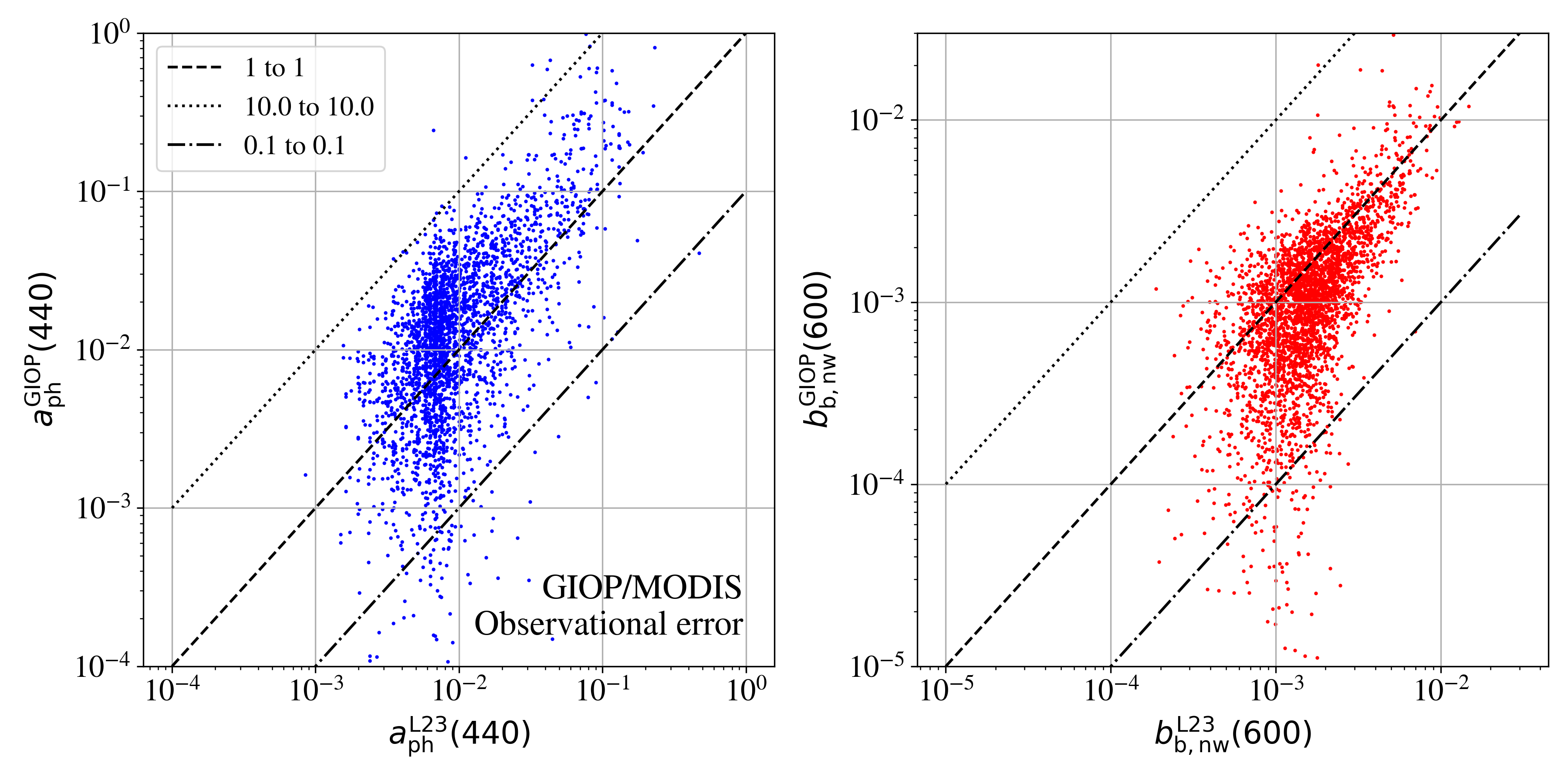}
\caption{Retrievals of \ffaph\ and \nbbnw\ 
at 600\,nm for the GIOP model with simulated
MODIS spectra and perturbing the \reflect\ 
values by the typical noise.
We find the values scatter by an order of 
magnitude or more (and have comparable 
uncertainties; next section).
}
\label{fig:GIOP_noise}
\end{figure}

\begin{figure}[ht!]
\centering\includegraphics[width=13cm]{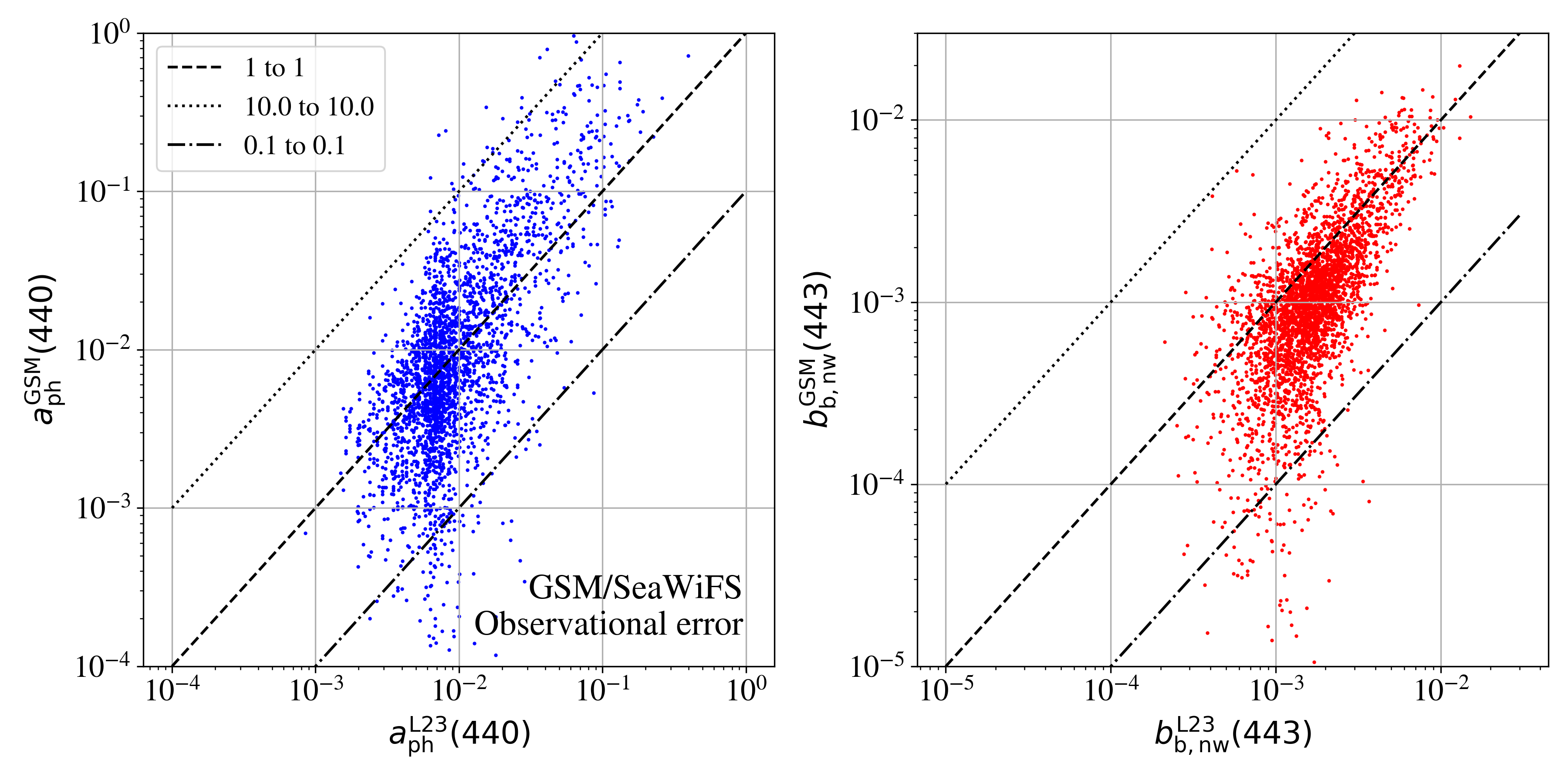}
\caption{Same as Figure~\ref{fig:GIOP_noise}
but for the GSM model and simulated \seawifs\ 
spectra.
}
\label{fig:GSM_noise}
\end{figure}

\begin{figure}[ht!]
\centering\includegraphics[width=13cm]{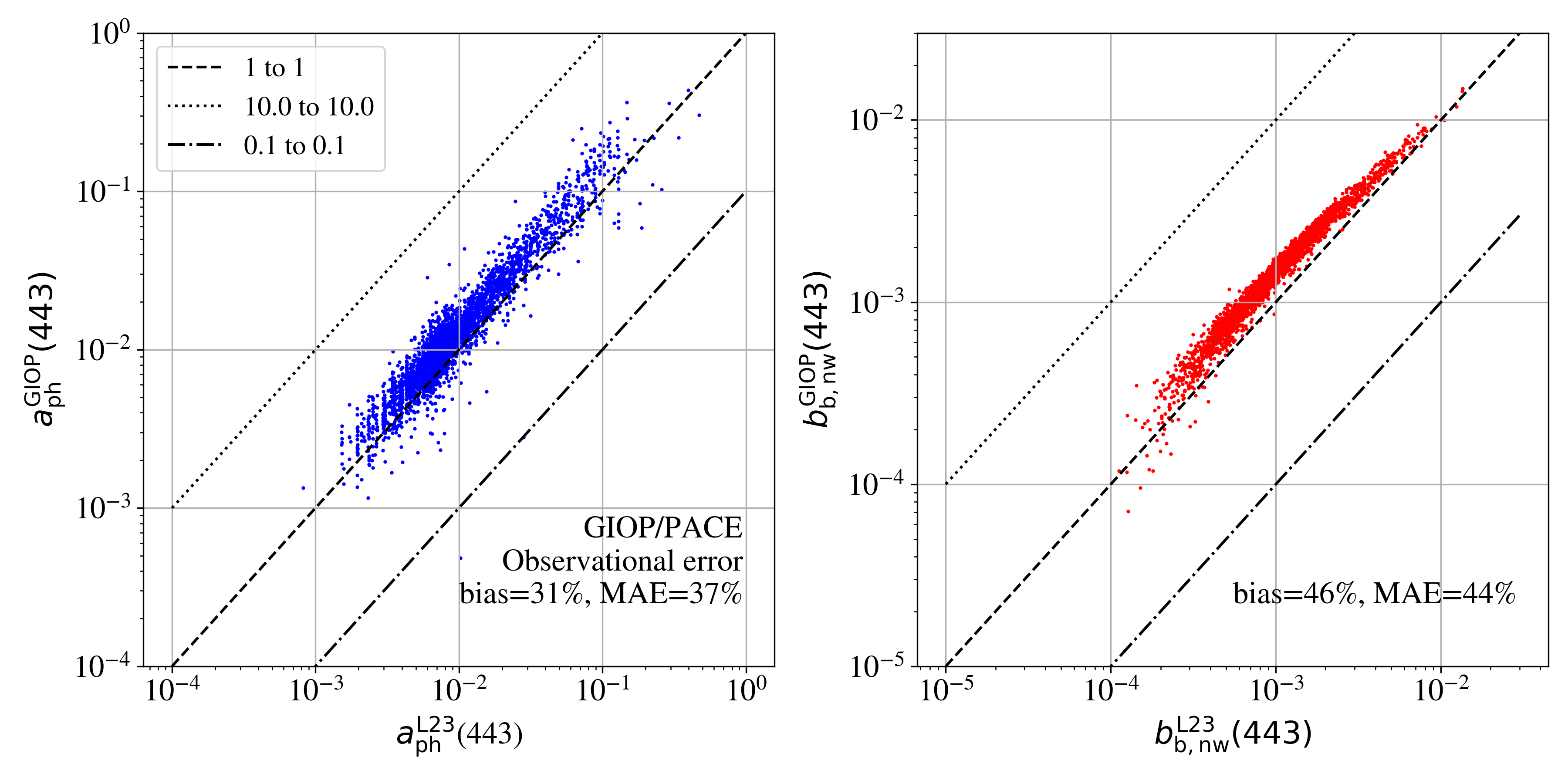}
\caption{Same as Figure~\ref{fig:GIOP_noise}
but for simulated PACE spectra
spectra.
We note that the mean absolute errors
(MAEs) estimated here for \naph\ and 
\nbb\ do meet the
uncertainty requirements listed in the final 
PACE technical report \cite{PACE_tech_vol6}.
}
\label{fig:GIOP_PACE_noise}
\end{figure}

\subsection{\aph\ Retrievals from GSM and GIOP}
\label{supp:gg}

While the results in Figure~\ref{fig:BIC_GIOP} and \ref{fig:BIC_GSM} lend
support that the \gsm\ and \giop\ models as designed may yield a positive 
detection of \aph, it is important to appreciate the limitations.
Consider first individual fits.  Figures~\ref{fig:GIOP_fit}
and \ref{fig:GSM_fit} show
\bing\ analyses on a representative spectrum from L23 (index=170) for 
GIOP with \modis\ observations and GSM on \seawifs.
It is evident that the algorithm yields only an upper limit in \bparam\ 
and that the estimates of \aparam\ and \aphparam\ are highly uncertain.
One notes the large uncertainties and the (weak) degeneracies between
the parameters.  Therefore, while one retrieves a non-zero, best-estimate
for \aph, \aphparam=10$^{-4}$\,nm$^{-1}$ is contained 
within the 90\%\ confidence interval.

We then performed a new set of inferences on the entire L23 dataset assuming \modis\ and \seawifs\ simulated spectra for the GIOP and GSM models
respectively.
In both cases, we calculated \reflect\ from Equations~\ref{eqn:rrs} and \ref{eqn:Rrs}
and performed the inversion with the same model and without perturbing
the \reflect\ values due to the presence of noise.
The retrievals are presented in Figures~\ref{fig:GIOP_perfect}
and \ref{fig:GSM_perfect}.
In the second, more realistic case, we perturbed the \reflect\ by the noise
model for each satellite (Tables~\ref{tab:seawifs}, \ref{tab:modis}),
and present these in Figures~\ref{fig:GIOP_noise} and \ref{fig:GSM_noise}.
Clearly, the retrievals of \naph\ and \nbbnw\ 
are highly uncertain at all values, with nearly
two orders of magnitude of scatter.
Therefore, the detection of \aph, if it were possible
with multi-spectral observations, would be
highly uncertain.

The constraints inherent in inversion algorithms like GSM and GIOP do affect the confidence in interpreting changes in maps of retrieved variables such as chlorophyll concentration, absorption coefficients, and backscattering coefficients. The spectral ambiguity in \reflect\ data 
can lead to changes influenced by variations in other optical properties not fully addressed by the models, making it difficult to attribute changes solely to biological factors. Moreover, the interdependence of retrieved parameters, such as chlorophyll, \acdom, and 
\nbbnw, means that errors in one can propagate to others, complicating the interpretation of these maps. 
For example, inaccuracies in backscatter coefficient estimates can affect chlorophyll retrievals. Additionally, variability in environmental conditions can impact the accuracy of the retrieved variables. Algorithm performance may vary across different 
water types and regions, necessitating further caution 
in interpreting these changes.

We have also performed retrievals with \giop\ on simulated \pace\ spectra
using the uncertainties described in Figure~\ref{fig:pace_noise}.
In contrast to the results from simulated \modis\ spectra
(Figure~\ref{fig:GIOP_noise}, the retrieved \ffaph\ and
$b_{\rm bp}(443)$ values have much reduced errors
(formally, $\approx 10\%)$.
However, each set of estimations is biased high by
several tens percent and the mean absolute error (MAE)
exceeds the \pace\ requirements \citep{PACE_tech_vol6}.
One could, of course, reduce the bias with one of several 
{\it post hoc} corrections (e.g.\ modify the \sparam\ value),
but this may have other, unanticipated consequences.
Furthermore, we reemphasize that our analysis assumes no errors
in the radiative transfer, and no uncertainties related to
spatial inhomogeneity (among other issues).
At present, we believe that \pace\ is not achieving one 
or more Level~1 scientific requirements \citep{PACE_tech_vol6}
and we are pessimistic that these may be met without including
strong and strict priors from other datasets in addition to
an entirely revamped methodology.


\subsection{Previous Assessments}
\label{supp:previous}

Results from tests of \iop\ retrieval models have been presented in multiple
publications over the years since the beginning of their development
\citep[e.g.][]{mouw+2017,werdell2018,seegers+2018}, and a discussion
of all of these lies beyond the scope of this manuscript.
Nevertheless, we wish to highlight one, in-depth effort summarized by
the IOCCG Report~5 \citep{IOCCG2006}.
Similar to our work, the participants applied their \iop\ retrieval algorithms
to a simulated (i.e.\ known) dataset to assess performance.
The majority of these algorithms assumed an exponential term for
\cdom/detritus absorption with fixed \sparam\ and a steep value
($\sparam > 0.015\,\spunit$).
Similar to the results we found for GIOP and GSM 
(Figure~\ref{fig:GSM_perfect},\ref{fig:GIOP_perfect}),
these consistently over-estimated \aph\ at 440\,nm.

Only one team (Boss \&\ Roesler) allowed 
\sparam\ to vary (from 0.008 to 0.023\,\spunit) in their 
algorithm which followed from the \cite{roesler1989} publication.  
Referring to their Figure~8.1, one notes less biased \aph\ 
values than the other algorithms and that they were the only group
to include an error estimation.
Given the axes are a log-scaling, one might miss that the 
uncertainties in \aph\ are large enough to be consistent with zero.
In other words, the results from the only algorithm of the
report adopting \sparam\ as a free parameter
showed \aph\ could not be positively detected from simulated data.

\subsection{New Methods for Hyperspectral Data}
\label{supp:new}

In the context of advancing ocean color remote sensing and PACE, 
new algorithms have been introduced 
to enhance the retrieval of phytoplankton pigments using hyperspectral reflectance data \citep{chase+2017,kramer2022}.
However, these approaches face significant challenges due to the fundamental issue of 
\reflect\ depending primarily on the ratio of backscattering to absorption (\nbb/\nabs) rather than on these parameters individually as emphasize
in the present study. This intrinsic degeneracy means that remote sensing reflectance alone cannot independently determine the absolute values of 
arbitrary non-water IOPs.

\cite{chase+2017}
employ an inversion algorithm to estimate phytoplankton pigments by using the spectral shape of reflectance. Their method relies heavily on covariation relationships between chlorophyll-a and accessory pigments, which does not resolve the degeneracy problem but instead depends on 
pre-established statistical correlations. Furthermore, their analysis of spectral residuals shows no strong relationship between residuals and pigments, underscoring the difficulty of extracting specific pigment information from hyperspectral reflectance alone.
Related, our own experiments implementing Gaussian fits to 
absorption spectra reveal major, insurmountable degeneracies
between the components \citep{bing_doi}.

Similarly, \cite{kramer2022} use principal components regression (PCR) modeling of the second derivative of \reflect\ residuals
to reconstruct pigment concentrations from 
hyperspectral  data. 
While they claim success in modeling multiple pigments, their approach remains fundamentally limited by the same degeneracy issue. The PCR method reduces dimensionality and identifies patterns in the data, yet it still relies on the representativeness of the training dataset and may not generalize well across diverse environmental conditions. Moreover, the requirement for high spectral resolution indicates that the model's robustness is contingent on specific sensor capabilities and noise levels, further limiting practical applications without extensive a priori data.
Indeed, our experiments on derivatives of \reflect\ reveal the 
uncertainties of the second derivative 
generally exceed the signal at all wavelengths
\citep{bing_doi}.

These studies, despite their contributions, illustrate the inherent limitations of using \reflect\ for detailed pigment retrieval without strong a priori knowledge or assumptions. Our analysis underscores the necessity of such priors to achieve accurate IOP retrievals, as 
\reflect\ alone cannot provide unique solutions due to its dependence on the $\nbb/\nabs$ ratio. Therefore, while the advancements by 
\cite{chase+2017} and \cite{kramer2022}
are noteworthy, their methodologies should be viewed within the context of these fundamental constraints, highlighting the need for continued refinement and incorporation of comprehensive 
a priori information in ocean color remote sensing algorithms.

\acknowledgments

The authors wish to thank B. M\'enard, E. Boss,
S. Kramer, A. Windle, H. Housekeeper, M. Zhang, and C. Mobley
for discussions and/or their input on an earlier draft.
We also benefited from conversations with The Astros.
Last, we acknowledge deriving new insight on the problem
from conversations with M. Kehrli.


%
\bibliography{bing}
%




%
%
%
%
%

\end{document}